# Theta*: Any-Angle Path Planning on Grids

**Kenny Daniel**                                                                  KFDANIEL@USC.EDU
**Alex Nash**                                                                         ANASH@USC.EDU
**Sven Koenig**                                                                    SKOENIG@USC.EDU
*Computer Science Department*
*University of Southern California*
*Los Angeles, California 90089-0781, USA*

**Ariel Felner**                                                                   FELNER@BGU.AC.IL
*Department of Information Systems Engineering*
*Ben-Gurion University of the Negev*
*Beer-Sheva, 85104, Israel*

## Abstract

Grids with blocked and unblocked cells are often used to represent terrain in robotics and video games. However, paths formed by grid edges can be longer than true shortest paths in the terrain since their headings are artificially constrained. We present two new correct and complete any-angle path-planning algorithms that avoid this shortcoming. Basic Theta* and Angle-Propagation Theta* are both variants of A* that propagate information along grid edges without constraining paths to grid edges. Basic Theta* is simple to understand and implement, fast and finds short paths. However, it is not guaranteed to find true shortest paths. Angle-Propagation Theta* achieves a better worst-case complexity per vertex expansion than Basic Theta* by propagating angle ranges when it expands vertices, but is more complex, not as fast and finds slightly longer paths. We refer to Basic Theta* and Angle-Propagation Theta* collectively as Theta*. Theta* has unique properties, which we analyze in detail. We show experimentally that it finds shorter paths than both A* with post-smoothed paths and Field D* (the only other version of A* we know of that propagates information along grid edges without constraining paths to grid edges) with a runtime comparable to that of A* on grids. Finally, we extend Theta* to grids that contain unblocked cells with non-uniform traversal costs and introduce variants of Theta* which provide different tradeoffs between path length and runtime.

## 1. Introduction

In this article, we study path planning for robotics and video games (Choset, Lynch, Hutchinson, Kantor, Burgard, Kavraki, & Thrun, 2005; Deloura, 2000; Patel, 2000; Murphy, 2000; Rabin, 2002), where a two-dimensional continuous terrain is discretized into a grid with blocked and unblocked cells. Our objective is to find a short unblocked path from a given start vertex to a given goal vertex (both at the corners of cells). A* finds grid paths (that is, paths constrained to grid edges) quickly, but grid paths are often not true shortest paths (that is, shortest paths in the terrain) since their potential headings are artificially constrained to multiples of 45 degrees, as shown in Figure 1(a) (Yap, 2002). This shortcoming led to the introduction of what we call any-angle path planning (Nash, Daniel, Koenig, & Felner, 2007; Ferguson & Stentz, 2006). Any-angle path-planning algorithms find paths without constraining the headings of the paths, as shown in Figure 1(b). We present two new correct and complete any-angle path-planning algorithms. Basic Theta* and Angle-Propagation Theta* are both variants of A* that propagate information along grid edges (to achieve a short runtime) without constraining paths to grid edges (to find any-angle paths). Unlike A* on visibility graphs, they are not guaranteed to find true shortest paths. The asterisk in their names thus does not denote their optimality but rather their similarity to A*. Basic Theta* is simple to understand and implement, fast and finds short paths. Angle-Propagation Theta* achieves a worst-case complexity per vertex expansion that is constant rather than linear in the number of cells (like that of Basic Theta*) by propagating angle ranges when it expands vertices, but is more complex, is not as fast and finds slightly longer paths. We refer to Basic Theta* and Angle-Propagation Theta* collectively as Theta*. Theta* has unique properties, which we analyze in detail. We show experimentally that it finds shorter paths than both A* with post-smoothed paths and Field D* (the only other version of A* we know of that propagates information along grid edges without constraining paths to grid edges) with a runtime comparable to that of A* on grids. Finally, we extend Theta* to grids that contain unblocked cells with non-uniform traversal costs and introduce variants of Theta* which provide different tradeoffs between path length and runtime.

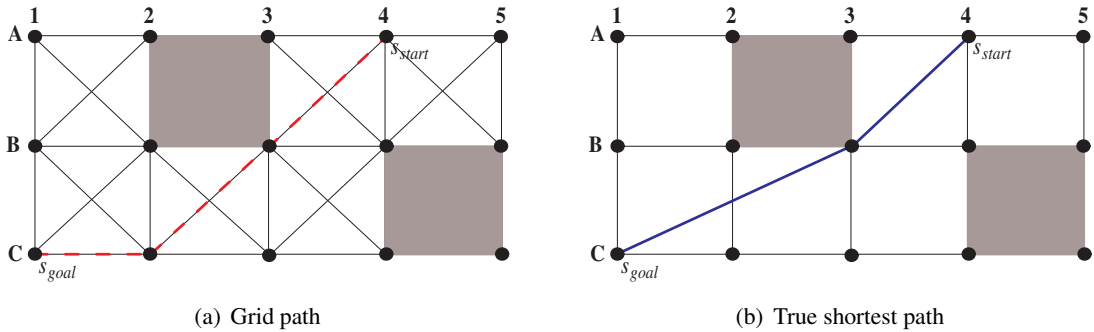

(a) Grid path          (b) True shortest path

Figure 1: Grid path versus true shortest path

## 2. Path-Planning Problem and Notation

In this section, we describe the path-planning problem that we study in this article, namely path planning on eight-neighbor grids with blocked and unblocked cells of uniform size. Cells are labeled as either blocked (grey) or unblocked (white). We use the corners of cells (rather than their centers) as vertices. $S$ is the set of all vertices. The path-planning problem is to find an unblocked path from a given start vertex $s_{start}$ to a given goal vertex $s_{goal}$.

A path is unblocked iff each vertex on the path has line-of-sight to its successor on the path. Vertex $s$ has line-of-sight to vertex $s'$, written as *LineOfSight*$(s, s')$, iff the straight line from vertex $s$ to vertex $s'$ neither passes through the interior of blocked cells nor passes between blocked cells that share an edge. Pseudocode for implementing the line-of-sight function is given in Appendix A. For simplicity, we allow a straight line to pass between diagonally touching blocked cells.

$c(s, s')$ is the length of the straight line from vertex $s$ to vertex $s'$. $nghbrs_{vis}(s)$ is the set of visible neighbors of vertex $s$ in the eight compass directions, that is those neighbors of vertex $s$ that have line-of-sight to vertex $s$. Figure 1 shows an example where the visible neighbors of vertex B4 are vertices A3, A4, A5, B3, B5, C3 and C4.

## 3. Existing Terrain Discretizations

Continuous terrain needs to be discretized for path planning. In this section, we compare grids to other existing terrain discretizations. We use grids to discretize terrain since they are widely used in robotics and video games (Deloura, 2000; Murphy, 2000; Rabin, 2004) and have several desirable properties:

- Grids are simple data structures and allow for simple path-planning algorithms.

- Terrain can easily be discretized into a grid by laying the grid over the terrain and labeling all cells that are partially or completely obstructed as blocked.

- Grids provide a comprehensive picture of all the traversable surfaces in the continuous terrain. This is essential when the path planning algorithm is used in a dynamic environment and must interact with a navigation planner. For example if a robot or video game character encounters a temporary blockage to its path, it can easily determine whether it is best to divert left (unblocked) or right (blocked) (Tozour, 2004).

- Cells can store information in addition to their traversability, such as the amount of gold hidden in the region of the terrain that corresponds to the cell or a rendering of the region when displaying the terrain.

- The information stored in cells can be accessed quickly since grids are random access data structures.

- The precision of path and navigation planning can be improved by simply increasing the grid resolution.

We now list some alternative terrain discretizations, assuming for simplicity that the obstacles in the terrain are polygonal.

- Voronoi graphs (Aurenhammer, 1991) discretize the terrain by biasing paths away from blocked polygons. The resulting paths can thus be much longer than true shortest paths.

- The discretization in the work of Mitchell and Papadimitriou (1991) partitions the terrain into regions with linear and hyperbolic edges, which allows one to find true shortest paths with time and space complexity $O(m^{5/3})$, where $m$ is the number of corners of blocked polygons. Thus, the runtime of path planning can grow superlinearly in the number of corners of blocked polygons.

- Framed Quadtrees (Yahja, Stentz, Singh, & Brumitt, 1998) recursively subdivide terrain into four equally sized cells until all cells are completely obstructed, completely unobstructed or of sufficiently small size. The resulting paths can have unnecessary heading changes (that is, heading changes that occur in free space rather than the corners of blocked polygons).

```
1  Main()
2      g(s_start) := 0;
3      parent(s_start) := s_start;
4      open := ∅;
5      open.Insert(s_start, g(s_start) + h(s_start));
6      closed := ∅;
7      while open ≠ ∅ do
8          s := open.Pop();
9          if s = s_goal then
10             return "path found";
11         closed := closed ∪ {s};
12         /* The following line is executed only by AP Theta*.  */;
13         [UpdateBounds(s)];
14         foreach s' ∈ nghbrs_vis(s) do
15             if s' ∉ closed then
16                 if s' ∉ open then
17                     g(s') := ∞;
18                     parent(s') := NULL;
19                 UpdateVertex(s, s');
20     return "no path found";
21 end
22 UpdateVertex(s,s')
23     if g(s) + c(s, s') < g(s') then
24         g(s') := g(s) + c(s, s');
25         parent(s') := s;
26         if s' ∈ open then
27             open.Remove(s');
28         open.Insert(s', g(s') + h(s'));
29 end
```

**Algorithm 1**: A*

- Probabilistic roadmaps (Kavraki, Svestka, Latombe, & Overmars, 1996) or rapidly-exploring random trees (LaValle & Kuffner, 2001) place vertices randomly (in addition to the start and goal vertex). Two vertices are connected via a straight line iff they have line-of-sight. The random placement of vertices needs to be tuned carefully since it influences the runtime of path planning, the likelihood of finding a path and the length of the path.

- Visibility graphs (Lee, 1978; Lozano-Pérez & Wesley, 1979) use the corners of each blocked polygon as vertices (in addition to the start and goal vertex). Two vertices are connected via a straight line iff they have line-of-sight, which allows one to find true shortest paths. The runtime of path planning can grow superlinearly in the number of vertices since the number of edges can grow quadratically in the number of vertices.

## 4. Existing Path-Planning Algorithms

In this section, we describe some existing path-planning algorithms, all of which are variants of A* (Hart, Nilsson, & Raphael, 1968). A* is a popular path-planning algorithm in robotics and video games. Algorithm 1 shows the pseudocode of A*. Line 13 is to be ignored. A* maintains three values for every vertex $s$:

- The g-value $g(s)$ is the length of the shortest path from the start vertex to vertex $s$ found so far and thus is an estimate of the start distance of vertex $s$.

- The user-provided h-value $h(s)$ is an estimate of the goal distance of vertex $s$. A* uses the h-value to calculate an f-value to focus the A* search. The f-value $f(s) = g(s) + h(s)$ is an estimate of the length of a shortest path from the start vertex via vertex $s$ to the goal vertex.

- The parent $parent(s)$ is used to extract a path from the start vertex to the goal vertex after A* terminates.

A* also maintains two global data structures:

- The open list is a priority queue that contains the vertices that A* considers for expansion. In the pseudocode, $open.Insert(s, x)$ inserts vertex $s$ with key $x$ into the priority queue $open$, $open.Remove(s)$ removes vertex $s$ from the priority queue $open$, and $open.Pop()$ removes a vertex with the smallest key from the priority queue $open$ and returns it.

- The closed list is a set that contains the vertices that A* has already expanded. It ensures that A* expands every vertex at most once.

A* sets the g-value of every vertex to infinity and the parent of every vertex to NULL when it encounters the vertex for the first time [Lines 17-18]. It sets the g-value of the start vertex to zero and the parent of the start vertex to the start vertex itself [Lines 2-3]. It sets the open and closed lists to the empty list and then inserts the start vertex into the open list with the f-value as its key [4-6]. A* then repeatedly executes the following procedure: If the open list is empty, then it reports that there is no path [Line 20]. Otherwise, it identifies a vertex $s$ with the smallest f-value in the open list [Line 8]. If this vertex is the goal vertex, then A* reports that it has found a path [Line 10]. Path extraction [not shown in the pseudocode] follows the parents from the goal vertex to the start vertex to retrieve a path from the start vertex to the goal vertex in reverse. Otherwise, A* removes the vertex from the open list [Line 8] and expands it by inserting the vertex into the closed list [Line 11] and then generating each of its unexpanded visible neighbors, as follows: A* checks whether the g-value of vertex $s$ plus the length of the straight line from vertex $s$ to vertex $s'$ is smaller than the g-value of vertex $s'$ [Line 23]. If so, then it sets the g-value of vertex $s'$ to the g-value of vertex $s$ plus the length of the straight line from vertex $s$ to vertex $s'$, sets the parent of vertex $s'$ to vertex $s$ and finally inserts vertex $s'$ into the open list with the f-value as its key or, if it was already in the open list, sets its key to the f-value [Lines 24-28]. It then repeats this procedure.

To summarize, when A* updates the g-value and parent of an unexpanded visible neighbor $s'$ of vertex $s$ in procedure UpdateVertex, it considers the path from the start vertex to vertex $s$ [= $g(s)$] and from vertex $s$ to vertex $s'$ in a straight line [= $c(s, s')$], resulting in a length of $g(s) + c(s, s')$ [Line 23]. A* updates the g-value and parent of vertex $s'$ if the considered path is shorter than the shortest path from the start vertex to vertex $s'$ found so far [= $g(s')$].

We now describe several existing path-planning algorithms that are versions of A* and how they trade off between two conflicting criteria, namely runtime and path length, as shown in Figure 2. We introduce them in order of decreasing path lengths.

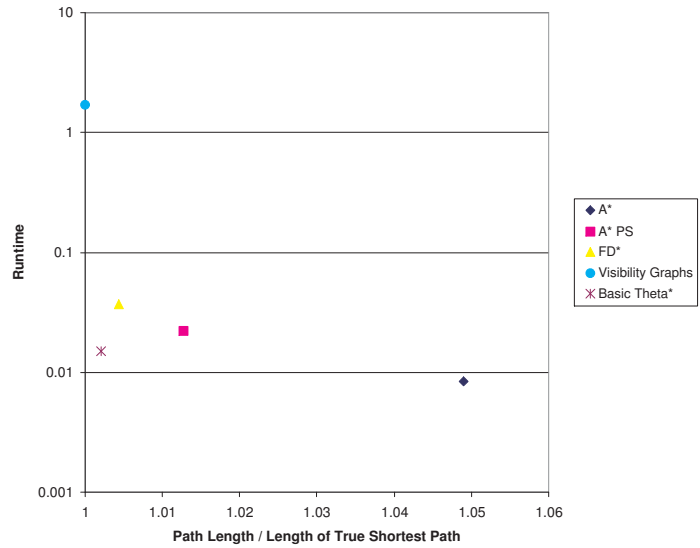

Figure 2: Runtime versus path length (relative to the length of true shortest path) on random $100 \times 100$ grids with 20 percent blocked cells

```
30  PostSmoothPath([s₀, . . . , sₙ])
31      k := 0;
32      tₖ := s₀;
33      foreach i := 1 . . . n − 1 do
34          if NOT LineOfSight(tₖ, sᵢ₊₁) then
35              k := k + 1;
36              tₖ := sᵢ;
37      k := k + 1;
38      tₖ := sₙ;
39      return [t₀, . . . , tₖ];
40  end
```

**Algorithm 2**: Post-smoothing

## 4.1 A* on Grids

One can run A* on grids, that is, on the graphs given by the grid vertices and edges. The resulting paths are artificially constrained to be formed by the edges of the grid, which can be seen in Figure 1(a). As a result the paths found by A* on grids are not equivalent to the true shortest paths and are unrealistic looking since they either deviate substantially from the true shortest paths or have many more heading changes, which provides the motivation for smoothing them. We use the octile distances, which can be computed using Algorithm 5, as h-values in the experiments.

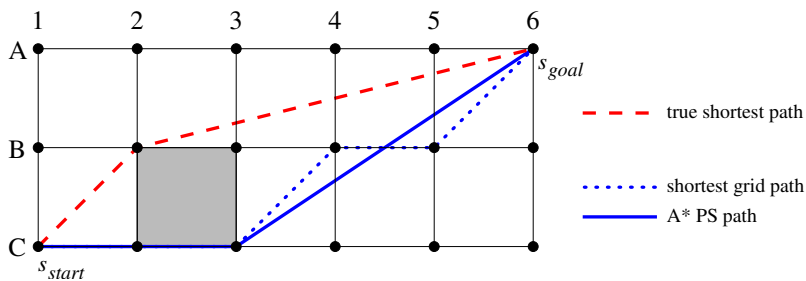

Figure 3: A* PS path versus true shortest path

## 4.2 A* with Post-Smoothed Paths (A* PS)

One can run A* with post-smoothed paths (A* PS) (Thorpe, 1984). A* PS runs A* on grids and then smoothes the resulting path in a post-processing step, which often shortens it at an increase in runtime. Algorithm 2 shows the pseudocode of the simple smoothing algorithm that A* PS uses in our experiments (Botea, Müller, & Schaeffer, 2004), which provides a good tradeoff between runtime and path length. Assume that A* on grids finds the path $[s_0, s_1, \ldots, s_n]$ with $s_0 = s_{start}$ and $s_n = s_{goal}$. A* PS uses the first vertex on the path as the current vertex. It then checks whether the current vertex $s_0$ has line-of-sight to the successor $s_2$ of its successor on the path. If so, A* PS removes the intermediate vertex $s_1$ from the path, thus shortening it. A* PS then repeats this procedure by checking again whether the current vertex $s_0$ has line-of-sight to the successor $s_3$ of its successor on the path, and so on. As soon as the current vertex does not have line-of-sight to the successor of its successor on the path, A* PS advances the current vertex and repeats this procedure until it reaches the end of the path. We use the straight-line distances $h(s) = c(s, s_{goal})$ as h-values in the experiments.

A* PS typically finds shorter paths than A* on grids, but is not guaranteed to find true shortest paths. Figure 3 shows an example. Assume that A* PS finds the dotted blue path, which is one of many shortest grid paths. It then smoothes this path to the solid blue path, which is not a true shortest path. The dashed red path, which moves above (rather than below) blocked cell B2-B3-C3-C2 is a true shortest path. A* PS is not guaranteed to find true shortest paths because it only considers grid paths during the A* search and thus cannot make informed decisions regarding other paths during the A* search, which motivates interleaving searching and smoothing. In fact, Theta* is similar to A* PS except that it interleaves searching and smoothing.

## 4.3 Field D* (FD*)

One can run Field D* (Ferguson & Stentz, 2006) (FD*). FD* propagates information along grid edges without constraining the paths to grid edges. FD* was designed to use D* Lite (Koenig & Likhachev, 2002) for fast replanning (by reusing information from the previous A* search to speed up the next one) and searches from the goal vertex to the start vertex. Our version of FD* uses A* and searches from the start vertex to the goal vertex, like all other path-planning algorithms in this article, which allows us to compare them fairly, except for their replanning abilities. (Theta* is currently in the process of being extended for fast replanning in Nash, Koenig, & Likhachev, 2009.)

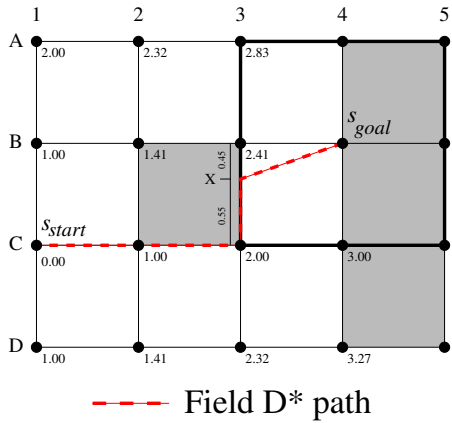

Figure 4: FD* path

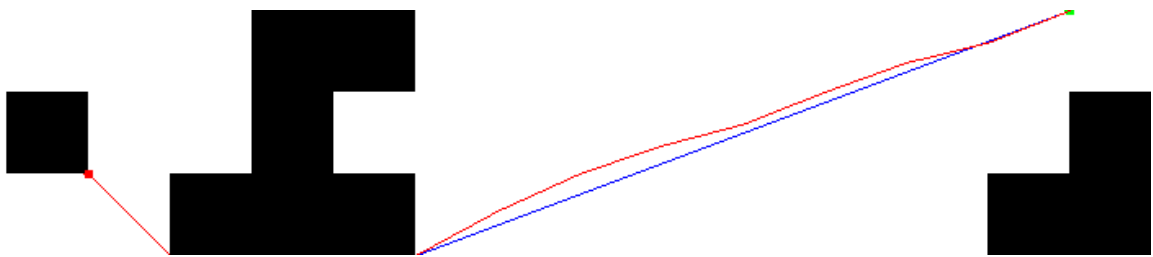

Figure 5: Screenshot of FD* path versus true shortest path

When FD* updates the g-value and parent of an unexpanded visible neighbor $s'$ of vertex $s$, it considers all paths from the start vertex to any point $X$ (not necessarily a vertex) on the perimeter of vertex $s'$ [= $g(X)$] that has line-of-sight to vertex $s'$, where the perimeter is formed by connecting all the neighbors of vertex $s'$, and from point $X$ to vertex $s'$ in a straight line [= $c(X, s')$], resulting in a length of $g(X) + c(X, s')$. FD* updates the g-value and parent of vertex $s'$ if the considered path is shorter than the shortest path from the start vertex to vertex $s'$ found so far [= $g(s')$]. We use the straight-line distances $h(s) = c(s, s_{goal})$ as h-values in the experiments.

Figure 4 shows an example. The perimeter of vertex $s' = B4$ is formed by connecting all of the neighbors of vertex B4, as shown in bold. Consider point $X$ on the perimeter. FD* does not know the g-value of point X since it only stores g-values for vertices. It calculates the g-value using linear interpolation between the g-values of the two vertices on the perimeter that are adjacent to the point X. Thus, it linearly interpolates between $g(B3) = 2.41$ and $g(C3) = 2.00$, resulting in $g(X) = 0.55 \times 2.41 + 0.45 \times 2.00 = 2.23$ since 0.55 and 0.45 are the distances from point $X$ to vertices B3 and C3, respectively. The calculated g-value of point $X$ is different from its true start distance [= 2.55] even though the g-values of vertices B3 and C3 are both equal to their true start distances. The reason for this mistake is simple. There exist true shortest paths from the start vertex through either vertex C3 or vertex B3 to the goal vertex. Thus, the linear interpolation assumption predicts that there must also exist a short path from the start vertex through any point along the edge that connects vertices B3 and C3 to the goal vertex. However, this is not the case since these

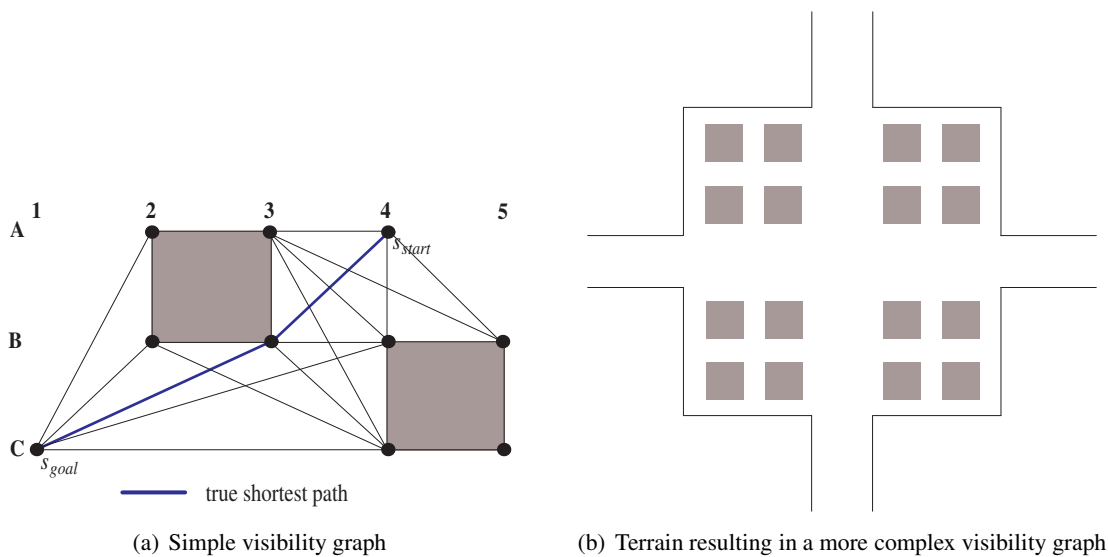

(a) Simple visibility graph        (b) Terrain resulting in a more complex visibility graph

Figure 6: Visibility graphs

paths need to circumnavigate blocked cell B2-B3-C3-C2, which makes them longer than expected. As a result of miscalculating the g-value of point $X$, FD* sets the parent of vertex $B4$ to point $X$, resulting in a path that has an unnecessary heading change at point $X$ and is longer than even a shortest grid path.

The authors of FD* recognize that the paths found by FD* frequently have unnecessary heading changes and suggest to use a one-step look-ahead algorithm during path extraction (Ferguson & Stentz, 2006), which FD* uses in our experiments. This one-step look-ahead algorithm allows FD* to avoid some of the unnecessary heading changes, like the one in Figure 4, but does not eliminate all of them. Figure 5 shows an example of an FD* path in red and the corresponding true shortest path in blue. The FD* path still has many unnecessary heading changes.

### 4.4 A* on Visibility Graphs

One can run A* on visibility graphs. The visibility graph of a grid with blocked and unblocked cells contains the start vertex, the goal vertex and the corners of all blocked cells (Lozano-Pérez & Wesley, 1979). We use the straight-line distances $h(s) = c(s, s_{goal})$ as h-values in the experiments. A* on visibility graphs finds true shortest paths, as shown in Figure 6(a). True shortest paths have heading changes only at the corners of blocked cells, while the paths found by A* on grids, A* PS and FD* can have unnecessary heading changes. On the other hand, A* on visibility graphs can be slow. It propagates information along visibility graph edges, whose number can grow quadratically in the number of cells, while A* on grids, A* PS and FD* propagate information along grid edges, whose number grows only linearly in the number of cells. If one constructed the visibility graphs before the A* search, one would need to perform a line-of-sight check for every pair of corners of blocked cells to determine whether or not there should be a visibility graph edge between them, which requires at least 2,556 line-of-sight checks for the room in Figure 6(b) (Tozour, 2004). The number of line-of-sight checks performed by A* on visibility graphs can be reduced by constructing

```
41  UpdateVertex(s,s')
42      if LineOfSight(parent(s), s') then
43          /* Path 2 */
44          if g(parent(s)) + c(parent(s), s') < g(s') then
45              g(s') := g(parent(s)) + c(parent(s), s');
46              parent(s') := parent(s);
47              if s' ∈ open then
48                  open.Remove(s');
49              open.Insert(s', g(s') + h(s'));
50      else
51          /* Path 1 */
52          if g(s) + c(s, s') < g(s') then
53              g(s') := g(s) + c(s, s');
54              parent(s') := s;
55              if s' ∈ open then
56                  open.Remove(s');
57              open.Insert(s', g(s') + h(s'));
58  end
```

**Algorithm 3**: Basic Theta*

the visibility graphs during the A* search. When it expands a vertex, it performs line-of-sight checks between the expanded vertex and the corners of all blocked cells (and the goal vertex). While this can significantly reduce the number of line-of-sight checks performed in some environments, such as simple outdoor terrain, it fails to do so in others, such as cluttered indoor terrain. More complex optimizations, such as reduced visibility graphs can further reduce the number of line-of-sight checks, but do not sufficiently speed up A* on visibility graphs (Liu & Arimoto, 1992).

## 5. Basic Theta*

In this section, we introduce Theta* (Nash et al., 2007), our version of A* for any-angle path planning that propagates information along grid edges without constraining the paths to grid edges. It combines the ideas behind A* on visibility graphs (where heading changes occur only at the corners of blocked cells) and A* on grids (where the number of edges grows only linearly in the number of cells). Its paths are only slightly longer than true shortest paths (as found by A* on visibility graphs), yet is only slightly slower than A* on grids, as shown in Figure 2. The key difference between Theta* and A* on grids is that the parent of a vertex can be any vertex when using Theta*, while the parent of a vertex has to be a neighbor of the vertex when using A*. We first introduce Basic Theta*, a simple version of Theta*.

Algorithm 3 shows the pseudocode of Basic Theta*. Procedure Main is identical to that of A* in Algorithm 1 and thus is not shown. Line 13 is to be ignored. We use the straight-line distances $h(s) = c(s, s_{goal})$ as h-values in the experiments.

### 5.1 Operation of Basic Theta*

Basic Theta* is simple. It is identical to A* except that, when it updates the g-value and parent of an unexpanded visible neighbor $s'$ of vertex $s$ in procedure UpdateVertex, it considers two paths instead of only the one path considered by A*. Figure 7(a) shows an example. Basic Theta* is expanding vertex B3 with parent A4 and needs to update the g-value and parent of unexpanded visible neighbor C3. Basic Theta* considers two paths:

- **Path 1:** Basic Theta* considers the path from the start vertex to vertex $s$ [$= g(s)$] and from vertex $s$ to vertex $s'$ in a straight line [$= c(s, s')$], resulting in a length of $g(s) + c(s, s')$ [Line 52]. Path 1 is the path considered by A*. It corresponds to the dashed red path [A4, B3, C3] in Figure 7(a).

- **Path 2:** Basic Theta* also considers the path from the start vertex to the parent of vertex $s$ [$= g(parent(s))$] and from the parent of vertex $s$ to vertex $s'$ in a straight line [$= c(parent(s), s')$], resulting in a length of $g(parent(s)) + c(parent(s), s')$ [Line 44]. Path 2 is not considered by A* and allows Basic Theta* to construct any-angle paths. It corresponds to the solid blue path [A4, C3] in Figure 7(a).

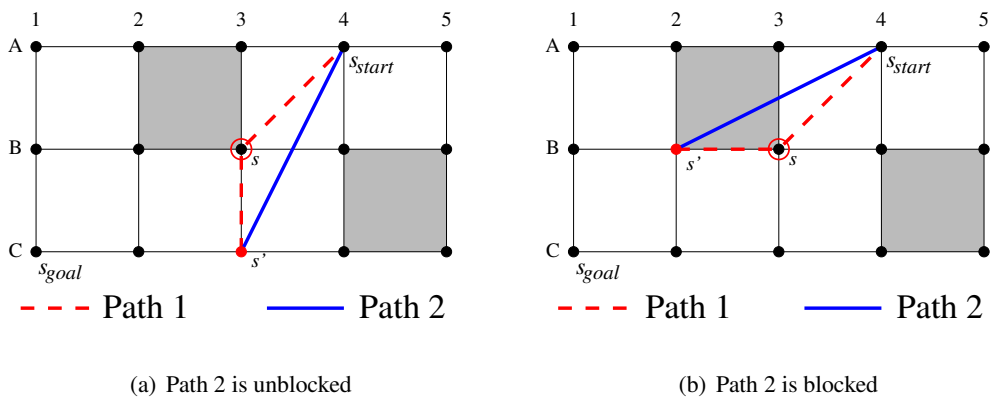

(a) Path 2 is unblocked       (b) Path 2 is blocked

Figure 7: Paths 1 and 2 considered by Basic Theta*

Path 2 is no longer than Path 1 due to the triangle inequality. The triangle inequality states that the length of any side of a triangle is no longer than the sum of the lengths of the other two sides. It applies here since Path 1 consists of the path from the start vertex to the parent of vertex $s$, the straight line from the parent of vertex $s$ to vertex $s$ (Line A) and the straight line from vertex $s$ to vertex $s'$ (Line B), Path 2 consists of the same path from the start vertex to the parent of vertex $s$ and the straight line from the parent of vertex $s$ to vertex $s'$ (Line C) and Lines A, B and C form a triangle. Path 1 is guaranteed to be unblocked but Path 2 is not. Thus, Basic Theta* chooses Path 2 over Path 1 if vertex $s'$ has line-of-sight to the parent of vertex $s$ and Path 2 is thus unblocked. Figure 7(a) shows an example. Otherwise, Basic Theta* chooses Path 1 over Path 2. Figure 7(b) shows an example. Basic Theta* updates the g-value and parent of vertex $s'$ if the chosen path is shorter than the shortest path from the start vertex to vertex $s'$ found so far [$= g(s')$]. We use the straight-line distances $h(s) = c(s, s_{goal})$ as h-values in the experiments.

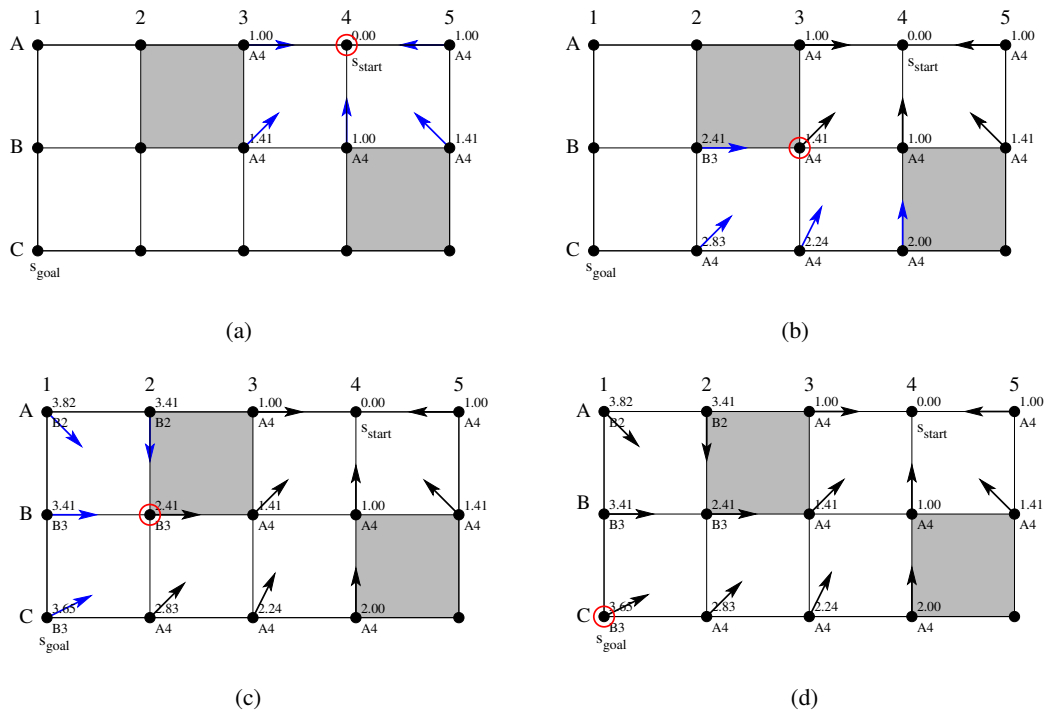

Figure 8: Example trace of Basic Theta*

## 5.2 Example Trace of Basic Theta*

Figure 8 shows an example trace of Basic Theta*. The vertices are labeled with their g-values and parents. The arrows point to their parents. Red circles indicate vertices that are being expanded, and blue arrows indicate vertices that are generated during the current expansion. First, Basic Theta* expands start vertex A4 with parent A4, as shown in Figure 8(a). It sets the parent of the unexpanded visible neighbors of vertex A4 to vertex A4, just like A* would do. Second, Basic Theta* expands vertex B3 with parent A4, as shown in Figure 8(b). Vertex B2 is an unexpanded visible neighbor of vertex B3 that does not have line-of-sight to vertex A4. Basic Theta* thus updates it according to Path 1 and sets its parent to vertex B3. On the other hand, vertices C2, C3 and C4 are unexpanded visible neighbors of vertex B3 that have line-of-sight to vertex A4. Basic Theta* thus updates them according to Path 2 and sets their parents to vertex A4. (The g-values and parents of the other unexpanded visible neighbors of vertex B3 are not updated.) Third, Basic Theta* expands vertex B2 with parent B3, as shown in Figure 8(c). Vertices A1 and A2 are unexpanded visible neighbors of vertex B2 that do not have line-of-sight to vertex B3. Basic Theta* thus updates them according to Path 1 and sets their parents to vertex B2. On the other hand, vertices B1 and C1 are unexpanded visible neighbors of vertex B2 that do have line-of-sight to vertex B3. Basic Theta* thus updates them according to Path 2 and sets their parents to vertex B3. Fourth, Basic Theta* expands goal vertex C1 with parent B3 and terminates, as shown in Figure 8(d). Path extraction then follows the parents from goal vertex C1 to start vertex A4 to retrieve the true shortest path [A4, B3, C1] from the start vertex to the goal vertex in reverse.

## 5.3 Properties of Basic Theta*

We now discuss the properties of Basic Theta*.

### 5.3.1 CORRECTNESS AND COMPLETENESS

Basic Theta* is correct (that is, finds only unblocked paths from the start vertex to the goal vertex) and complete (that is, finds a path from the start vertex to the goal vertex if one exists). We use the following lemmata in the proof.

**Lemma 1.** *If there exists an unblocked path between two vertices then there also exists an unblocked grid path between the same two vertices.*

*Proof.* An unblocked path between two vertices exists iff an unblocked any-angle path $[s_0, \ldots, s_n]$ exists between the same two vertices. Consider any path segment $\overline{s_k s_{k+1}}$ of this any-angle path. If the path segment is horizontal or vertical, then consider the unblocked grid path from vertex $s_k$ to vertex $s_{k+1}$ that coincides with the path segment. Otherwise, consider the sequence $(b_0, \ldots, b_m)$ of unblocked cells whose interior the path segment passes through. Any two consecutive cells $b_j$ and $b_{j+1}$ share at least one vertex $s'_{j+1}$ since the cells either share an edge or are diagonally touching. (If they share more than one vertex, pick one arbitrarily.) Consider the grid path $[s'_0 = s_k, s'_1, \ldots, s'_m, s'_{m+1} = s_{k+1}]$. This grid path from vertex $s_k$ to vertex $s_{k+1}$ is unblocked since any two consecutive vertices on it are corners of the same unblocked cell and are thus visible neighbors. Repeat this procedure for every path segment of the any-angle path and concatenate the resulting grid paths to an unblocked grid path from vertex $s_0$ to vertex $s_n$. (If several consecutive vertices on the grid path are identical, then all of them but one can be removed.) ☐

**Lemma 2.** *At any point during the execution of Basic Theta*, following the parents from any vertex in the open or closed lists to the start vertex retrieves an unblocked path from the start vertex to this vertex in reverse.*

*Proof.* We prove by induction that the lemma holds and that the parent of any vertex in the union of the open or closed lists itself is in the union of the open or closed lists. This statement holds initially because the start vertex is the only vertex in the union of the open or closed lists and it is its own parent. We now show that the statement continues to hold whenever a vertex changes either its parent or its membership in the union of the open or closed lists. Once a vertex is a member of the union of the open or closed lists, it continues to be a member. A vertex can become a member in the union of the open or closed lists only when Basic Theta* expands some vertex $s$ and updates the g-value and parent of an unexpanded visible neighbor $s'$ of vertex $s$ in procedure UpdateVertex. Vertex $s$ is thus in the closed list, and its parent is in the union of the open or closed lists according to the induction assumption. Thus, following the parents from vertex $s$ (or its parent) to the start vertex retrieves an unblocked path from the start vertex to vertex $s$ (or its parent, respectively) in reverse according to the induction assumption. If Basic Theta* updates vertex $s'$ according to Path 1, then the statement continues to hold since vertices $s$ and $s'$ are visible neighbors and the path segment from vertex $s$ to vertex $s'$ is thus unblocked. If Basic Theta* updates vertex $s'$ according to Path 2, then the statement continues to hold since Basic Theta* explicitly checks that the path segment from the parent of vertex $s$ to vertex $s'$ is unblocked. There are no other ways in which the parent of a vertex can change. □

**Theorem 1.** *Basic Theta\* terminates and path extraction retrieves an unblocked path from the start vertex to the goal vertex if such a path exists. Otherwise, Basic Theta\* terminates and reports that no unblocked path exists.*

*Proof.* The following properties together prove the theorem. Their proofs utilize the fact that Basic Theta\* terminates iff the open is empty or it expands the goal vertex. The start vertex is initially in the open list. Any other vertex is initially neither in the open nor closed lists. A vertex neither in the open nor closed lists can be inserted into the open list. A vertex in the open list can be removed from the open list and be inserted into the closed list. A vertex in the closed list remains in the closed list.

- Property 1: Basic Theta\* terminates. It expands one vertex in the open list during each iteration. In the process, it removes the vertex from the open list and can then never insert it into the open list again. Since the number of vertices is finite, the open list eventually becomes empty and Basic Theta\* has to terminate if it has not terminated earlier already.

- Property 2: If Basic Theta\* terminates because its open list is empty, then there does not exist an unblocked path from the start vertex to the goal vertex. We prove the contrapositive. Assume that there exists an unblocked path from the start vertex to the goal vertex. We prove by contradiction that Basic Theta\* then does not terminate because its open list is empty. Thus, assume also that Basic Theta\* terminates because its open list is empty. Then, there exists an unblocked grid path $[s_0 = s_{start}, \ldots, s_n = s_{goal}]$ from the start vertex to the goal vertex according to Lemma 1. Choose vertex $s_i$ to be the first vertex on the grid path that is not in the closed list when Basic Theta\* terminates. The goal vertex is not in the closed list when Basic Theta\* terminates since Basic Theta\* would otherwise have terminated when it expanded the goal vertex. Thus, vertex $s_i$ exists. Vertex $s_i$ is not the start vertex since the start vertex would otherwise be in the open list and Basic Theta\* could not have terminated because its open list is empty. Thus, vertex $s_i$ has a predecessor on the grid path. This predecessor is in the closed list when Basic Theta\* terminates since vertex $s_i$ is the first vertex on the grid path that is not in the closed list when Basic Theta\* terminates. When Basic Theta\* expanded the predecessor, it added vertex $s_i$ to the open list. Thus, vertex $s_i$ is still in the open list when Basic Theta\* terminates. But then Basic Theta\* could not have terminated because its open list is empty, which is a contradiction.

- Property 3: If Basic Theta\* terminates because it expands the goal vertex, then path extraction retrieves an unblocked path from the start vertex to the goal vertex because following the parents from the goal vertex to the start vertex retrieves an unblocked path from the start vertex to the goal vertex in reverse according to Lemma 2.

□

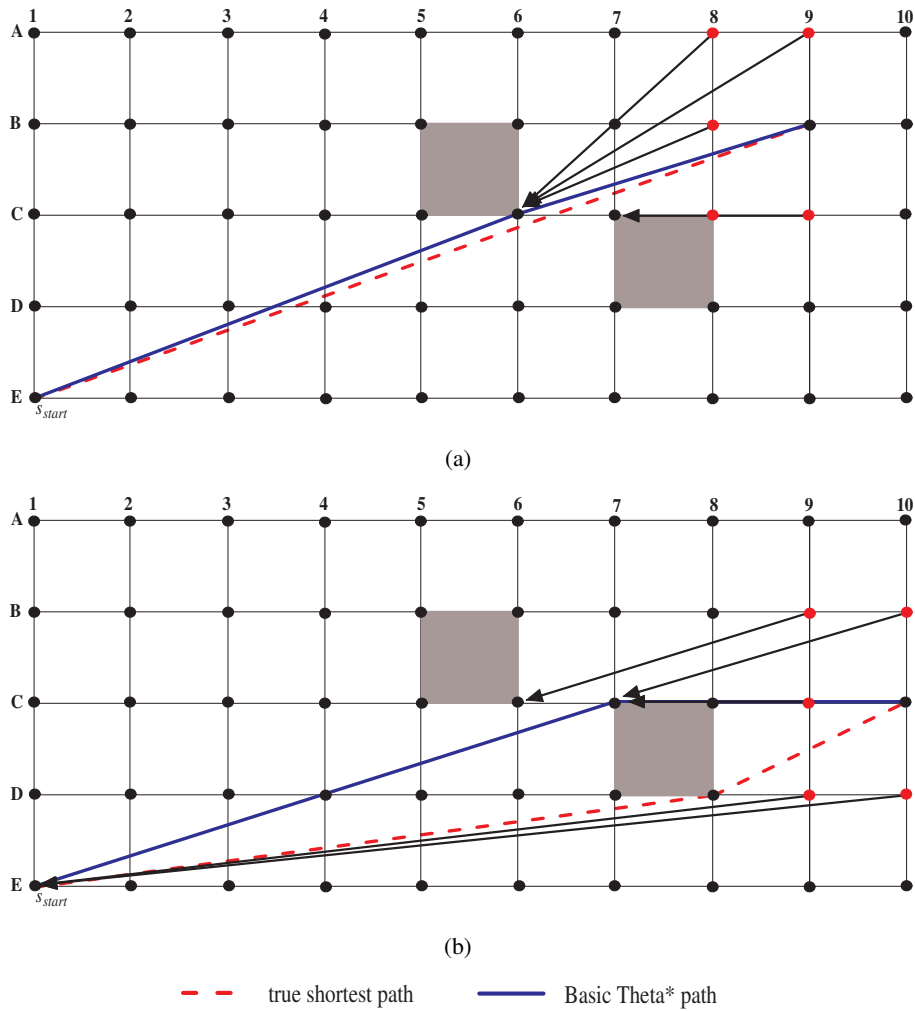

(a)

(b)

true shortest path — — —   Basic Theta* path ————

Figure 9: Basic Theta* paths versus true shortest paths

### 5.3.2 OPTIMALITY

Basic Theta* is not optimal (that is, it is not guaranteed to find true shortest paths) because the parent of a vertex has to be either a visible neighbor of the vertex or the parent of a visible neighbor, which is not always the case for true shortest paths. Figure 9(a) shows an example where the dashed red path [E1, B9] is a true shortest path from start vertex E1 to vertex B9 since vertex E1 has line-of-sight to vertex B9. However, vertex E1 is neither a visible neighbor nor the parent of a visible neighbor of vertex B9 since vertex E1 does not have line-of-sight to these vertices (highlighted in red). Thus, Basic Theta* cannot set the parent of vertex B9 to vertex E1 and does not find a true shortest path from vertex E1 to vertex B9. Similarly, Figure 9(b) shows an example where the dashed red path [E1, D8, C10] is a true shortest path from vertex E1 to vertex C10. However, vertex D8 is neither a visible neighbor nor the parent of a visible neighbor of vertex C10 since start vertex E1 either has line-of-sight to them or Basic Theta* found paths from vertex E1 to them that do not

547

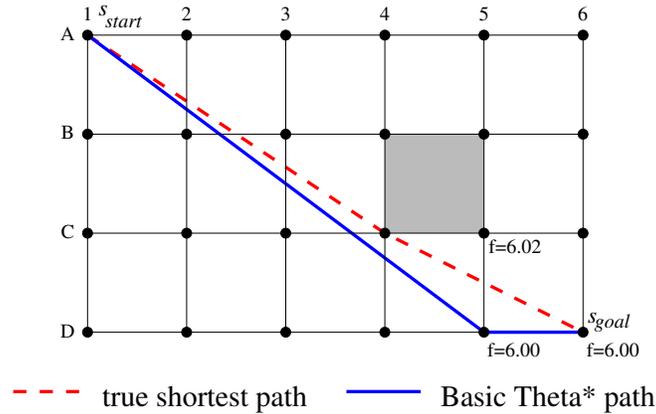

Figure 10: Heading changes of Basic Theta*

### 5.3.3 HEADING CHANGES

Basic Theta* takes advantage of the fact that true shortest paths have heading changes only at the corners of blocked cells. However, the paths found by Basic Theta* can occasionally have unnecessary heading changes. Figure 10 shows an example where Basic Theta* finds the solid blue path [A1, D5, D6] from vertex A1 to vertex D6. The reason for this mistake is simple. Assume that the open list contains both vertices C5 and D5. The f-value of vertex C5 is $f(C5) = g(C5) + h(C5) = 4.61 + 1.41 = 6.02$ and its parent is vertex C4. The f-value of vertex D5 is $f(D5) = 5.00 + 1.00 = 6.00$ and its parent is vertex A1. Thus Basic Theta* expands vertex D5 before vertex C5 (since its f-value is smaller). When Basic Theta* expands vertex D5 with parent A1, it generates vertex D6. Vertex D6 is an unexpanded visible neighbor of vertex D5 that does not have line-of-sight to vertex A1. Basic Theta* thus updates it according to Path 1, sets its f-value to $f(D6) = 6.00 + 0.00 = 6.00$ and inserts it into the open list. Thus Basic Theta* expands goal vertex D6 before vertex C5 (since its f-value is smaller) and terminates. Path extraction then follows the parents from goal vertex D6 to start vertex A1 to retrieve the solid blue path [A1, D5, D6]. Thus, Basic Theta* never expands vertex C5, which would have resulted in it setting the parent of vertex D6 to vertex C4 according to Path 2 and path extraction retrieving the dashed red path [A1, C4, D6] which is the true shortest path. The solid blue path from vertex A1 to vertex D6 in Figure 10 is less than a factor of 1.027 longer than true shortest path.

```
59  UpdateVertex(s,s')
60      if s ≠ s_start AND lb(s) ≤ Θ(s, parent(s), s') ≤ ub(s) then
61          /* Path 2 */
62          if g(parent(s)) + c(parent(s), s') < g(s') then
63              g(s') := g(parent(s)) + c(parent(s), s');
64              parent(s') := parent(s);
65              if s' ∈ open then
66                  open.Remove(s');
67              open.Insert(s', g(s') + h(s'));
68      else
69          /* Path 1 */
70          if g(s) + c(s, s') < g(s') then
71              g(s') := g(s) + c(s, s');
72              parent(s') := s;
73              if s' ∈ open then
74                  open.Remove(s');
75              open.Insert(s', g(s') + h(s'));
76  end
77  UpdateBounds(s)
78      lb(s) := -∞; ub(s) := ∞;
79      if s ≠ s_start then
80          foreach blocked cell b adjacent to s do
81              if ∀s' ∈ corners(b) : parent(s) = s' OR Θ(s, parent(s), s') < 0 OR
82              (Θ(s, parent(s), s') = 0 AND c(parent(s), s') ≤ c(parent(s), s)) then
83                  lb(s) = 0;
84              if ∀s' ∈ corners(b) : parent(s) = s' OR Θ(s, parent(s), s') > 0 OR
85              (Θ(s, parent(s), s') = 0 AND c(parent(s), s') ≤ c(parent(s), s)) then
86                  ub(s) = 0;
87          foreach s' ∈ nghbrs_vis(s) do
88              if s' ∈ closed AND parent(s) = parent(s') AND s' ≠ s_start then
89                  if lb(s') + Θ(s, parent(s), s') ≤ 0 then
90                      lb(s) := max(lb(s), lb(s') + Θ(s, parent(s), s'));
91                  if ub(s') + Θ(s, parent(s), s') ≥ 0 then
92                      ub(s) := min(ub(s), ub(s') + Θ(s, parent(s), s'));
93              if c(parent(s), s') < c(parent(s), s) AND parent(s) ≠ s' AND (s' ∉ closed OR parent(s) ≠ parent(s'))
                then
94                  if Θ(s, parent(s), s') < 0 then
95                      lb(s) := max(lb(s), Θ(s, parent(s), s'));
96                  if Θ(s, parent(s), s') > 0 then
97                      ub(s) := min(ub(s), Θ(s, parent(s), s'));
98  end
```

**Algorithm 4**: AP Theta*

## 6. Angle-Propagation Theta* (AP Theta*)

The runtime of Basic Theta* per vertex expansion (that is, the runtime consumed during the generation of the unexpanded visible neighbors when expanding a vertex) can be linear in the number of cells since the runtime of each line-of-sight check can be linear in the number of cells. In this section, we introduce Angle-Propagation Theta* (AP Theta*), which reduces the runtime of Basic

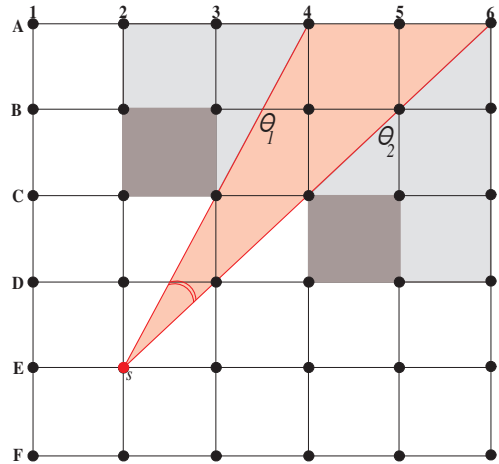

Figure 11: Region of points with line-of-sight to vertex $s$

Theta* per vertex expansion from linear to constant.[1] The key difference between AP Theta* and Basic Theta* is that AP Theta* propagates angle ranges and uses them to determine whether or not two vertices have line-of-sight.

If there is a light source at a vertex and light cannot pass through blocked cells, then cells in the shadows do not have line-of-sight to the vertex while all other cells have line-of-sight to the vertex. Each contiguous region of points that have line-of-sight to the vertex can be characterized by two rays emanating from the vertex and thus by an angle range defined by two angle bounds. Figure 11 shows an example where all points within the red angle range defined by the two angle bounds $\theta_1$ and $\theta_2$ have line-of-sight to vertex $s$. AP Theta* calculates the angle range of a vertex when it expands the vertex and then propagates it along grid edges, resulting in a constant runtime per vertex expansion since the angle ranges can be propagated in constant time and the line-of-sight checks can be performed in constant time as well.

Algorithm 4 shows the pseudocode of AP Theta*. Procedure Main is identical to that of A* in Algorithm 1 and thus is not shown. Line 13 is to be executed. We use the straight-line distances $h(s) = c(s, s_{goal})$ as h-values in the experiments.

## 6.1 Definition of Angle Ranges

We now discuss the key concept of an angle range. AP Theta* maintains two additional values for every vertex $s$, namely a lower angle bound $lb(s)$ of vertex $s$ and an upper angle bound $ub(s)$ of vertex $s$, that together form the angle range $[lb(s), ub(s)]$ of vertex $s$. The angle bounds correspond to headings of rays (measured in degrees) that originate at the parent of vertex $s$. The heading of the ray from the parent of vertex $s$ to vertex $s$ is zero degrees. A visible neighbor of vertex $s$ is guaranteed to have line-of-sight to the parent of vertex $s$ if (but not necessarily only if) the heading of the ray from the parent of vertex $s$ to the visible neighbor of vertex $s$ is contained in the angle

---

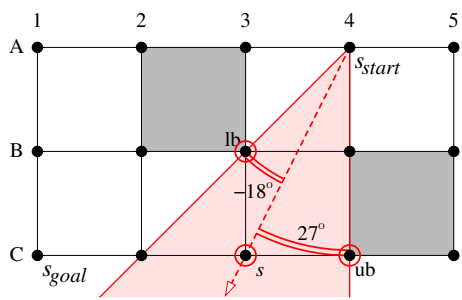

Figure 12: Angle range of AP Theta*

range of vertex $s$. Figure 12 shows an example where vertex C3 with parent A4 has angle range $[-18, 27]$. Thus, all visible neighbors of vertex C3 in the red region are guaranteed to have line-of-sight to the parent of vertex C3. For example, vertex C4 is guaranteed to have line-of-sight to the parent of vertex C3 but vertex B2 is not. AP Theta* therefore assumes that vertex B2 does not have line-of-sight to the parent of vertex C3.

We now define the concept of an angle range more formally. $\Theta(s, p, s') \in [-90, 90]$, which gives AP Theta* its name, is the angle (measured in degrees) between the ray from vertex $p$ to vertex $s$ and the ray from vertex $p$ to vertex $s'$. It is positive if the ray from vertex $p$ to vertex $s$ is clockwise from the ray from vertex $p$ to vertex $s'$, zero if the ray from vertex $p$ to vertex $s$ has the same heading as the ray from vertex $p$ to vertex $s'$, and negative if the ray from vertex $p$ to vertex $s$ is counterclockwise from the ray from vertex $p$ to vertex $s'$. Figure 12 shows an example where $\Theta(C3, A4, C4) = 27$ and $\Theta(C3, A4, B3) = -18$. A visible neighbor $s'$ of vertex $s$ is guaranteed to have line-of-sight to the parent of vertex $s$ if (but not necessarily only if) $lb(s) \leq \Theta(s, parent(s), s') \leq ub(s)$ (Visibility Property).

## 6.2 Update of Angle Ranges

We now discuss how AP Theta* calculates the angle range of a vertex when it expands the vertex. This calculation is complicated by the fact that AP Theta* is not guaranteed to have sufficient information to determine the angle range exactly since the order of vertex expansions depends on a variety of factors, such as the h-values. In this case, AP Theta* can constrain the angle range more than necessary to guarantee that the Visibility Property holds and that it finds unblocked paths.

When AP Theta* expands vertex $s$, it sets the angle range of vertex $s$ initially to $[-\infty, \infty]$, meaning that all visible neighbors of the vertex are guaranteed to have line-of-sight to the parent of the vertex. It then constrains the angle range more and more if vertex $s$ is not the start vertex.

AP Theta* constrains the angle range of vertex $s$ based on each blocked cell $b$ that is adjacent to vertex $s$ (that is, that vertex $s$ is a corner of $b$, written as $s \in corners(b)$) provided that at least one of two conditions is satisfied:

- **Case 1:** If every corner $s'$ of blocked cell $b$ satisfies at least one of the following conditions:

  - $parent(s) = s'$ or

    – $\Theta(s, parent(s), s') < 0$ or

    – $\Theta(s, parent(s), s') = 0$ and $c(parent(s), s') \leq c(parent(s), s)$,

then AP Theta* assumes that a vertex $s''$ does not have line-of-sight to the parent of vertex $s$ if the ray from the parent of vertex $s$ to vertex $s$ is counterclockwise from the ray from the parent of vertex $s$ to vertex $s''$, that is, if $\Theta(s, parent(s), s'') < 0$. AP Theta* therefore sets the lower angle bound of vertex $s$ to $\Theta(s, parent(s), s) = 0$ [Line 83].

- **Case 2:** If every corner $s'$ of blocked cell $b$ satisfies at least one of the following conditions:

  – $parent(s) = s'$ or

  – $\Theta(s, parent(s), s') > 0$ or

  – $\Theta(s, parent(s), s') = 0$ and $c(parent(s), s') \leq c(parent(s), s)$,

  then AP Theta* assumes that a vertex $s''$ does not have line-of-sight to the parent of vertex $s$ if the ray from the parent of vertex $s$ to vertex $s$ is clockwise from the ray from the parent of vertex $s$ to vertex $s''$, that is, if $\Theta(s, parent(s), s'') > 0$. AP Theta* therefore sets the upper angle bound of vertex $s$ to $\Theta(s, parent(s), s) = 0$ [Line 86].

AP Theta* also constrains the angle range of vertex $s$ based on each visible neighbor $s'$ of vertex $s$ provided that at least one of two conditions is satisfied:

- **Case 3:** If vertex $s'$ satisfies all of the following conditions:

  – $s' \in closed$ and

  – $parent(s) = parent(s')$ and

  – $s' \neq s_{start}$,

  then AP Theta* constrains the angle range of vertex $s$ by intersecting it with the angle range of vertex $s'$ [Lines 90 and 92]. To do that, it first shifts the angle range of vertex $s'$ by $\Theta(s, parent(s), s')$ degrees to take into account that the angle range of vertex $s'$ is calibrated so that the heading of the ray from the joint parent of vertices $s$ and $s'$ to vertex $s'$ is zero degrees, while the angle range of vertex $s$ is calibrated so that the heading of the ray from the joint parent of vertices $s$ and $s'$ to vertex $s$ is zero degrees. Lines 89 and 91 ensure that the lower angle bound always remains non-positive and the upper angle bound always remains non-negative, respectively. The fact that lower angle bounds should be non-positive (and upper angle bounds non-negative) is intuitive in that if a vertex $s$ is assigned parent vertex $p$ then the angle of the ray from vertex $p$ to vertex $s$ should be included in the angle range of vertex $s$.

- **Case 4:** If vertex $s'$ satisfies all of the following conditions:

  – $c(parent(s), s') < c(parent(s), s)$ and

  – $parent(s) \neq s'$ and

- $s' \notin closed$ or $parent(s) \neq parent(s')$,

then AP Theta* has insufficient information about vertex $s'$. AP Theta* therefore cannot determine the angle range of vertex $s$ exactly and makes the conservative assumption that vertex $s'$ barely has line-of-sight to the parent of vertex $s$ [Lines 95 and 97].

The Visibility Property holds after AP Theta* has updated the angle range of vertex $s$ in procedure UpdateBounds. Thus, when AP Theta* checks whether or not a visible neighbor $s'$ of vertex $s$ has line-of-sight to the parent of vertex $s$, it now checks whether or not $lb(s) \leq \Theta(s, parent(s), s') \leq ub(s)$ [Line 60] is true instead of whether or not $LineOfSight(parent(s), s')$ [Line 42] is true. These are the only differences between AP Theta* and Basic Theta*.

Figure 13(a) shows an example where AP Theta* calculates the angle range of vertex A4. It sets the angle range to $[-\infty, \infty]$. Figure 13(b) shows an example where AP Theta* calculates the angle range of vertex B3. It sets the angle range initially to $[-\infty, \infty]$. It then sets the lower angle bound to 0 degrees according to Case 1 based on the blocked cell A2-A3-B3-B2 [Line 83]. It sets the upper angle bound to 45 degrees according to Case 4 based on vertex B4, which is unexpanded and thus not in the closed list [Line 97]. Figure 13(c) shows an example where AP Theta* calculates the angle range of vertex B2. It sets the angle range initially to $[-\infty, \infty]$. It then sets the lower angle bound to 0 degrees according to Case 1 based on the blocked cell A2-A3-B3-B2 [Line 83]. Assume that vertex C1 is not the goal vertex. Figure 13(d) then shows an example where AP Theta* calculates the angle range of vertex C1. It sets the angle range initially to $[-\infty, \infty]$. It then sets the lower angle bound to -27 degrees according to Case 3 based on vertex B2 [Line 90] and the upper angle bound to 18 degrees according to Case 4 based on vertex C2, which is unexpanded and thus not in the closed list [Line 97].

### 6.3 Example Trace of AP Theta*

Figure 13 shows an example trace of AP Theta* using the path-planning problem from Figure 8. The labels of the vertices now include the angle ranges.

### 6.4 Properties of AP Theta*

We now discuss the properties of AP Theta*. AP Theta* operates in the same way as Basic Theta* and thus has similar properties as Basic Theta*. For example, AP Theta* is correct and complete. It is not guaranteed to find true shortest paths, and its paths can occasionally have unnecessary heading changes.

AP Theta* sometimes constrains the angle ranges more than necessary to guarantee that it finds unblocked paths, which means that its line-of-sight checks sometimes fail incorrectly in which case it has to update vertices according to Path 1 rather than Path 2. AP Theta* is still complete since it finds an unblocked grid path if all line-of-sight checks fail, and there always exists an unblocked grid path if there exists an unblocked any-angle path. However, the paths found by AP Theta* can be longer than those found by Basic Theta*. Figure 14 shows an example. When AP Theta* expands vertex C4 with parent B1 and calculates the angle range of vertex C4, vertex C3 is unexpanded and thus not in the closed list. This means that AP Theta* has insufficient information about vertex

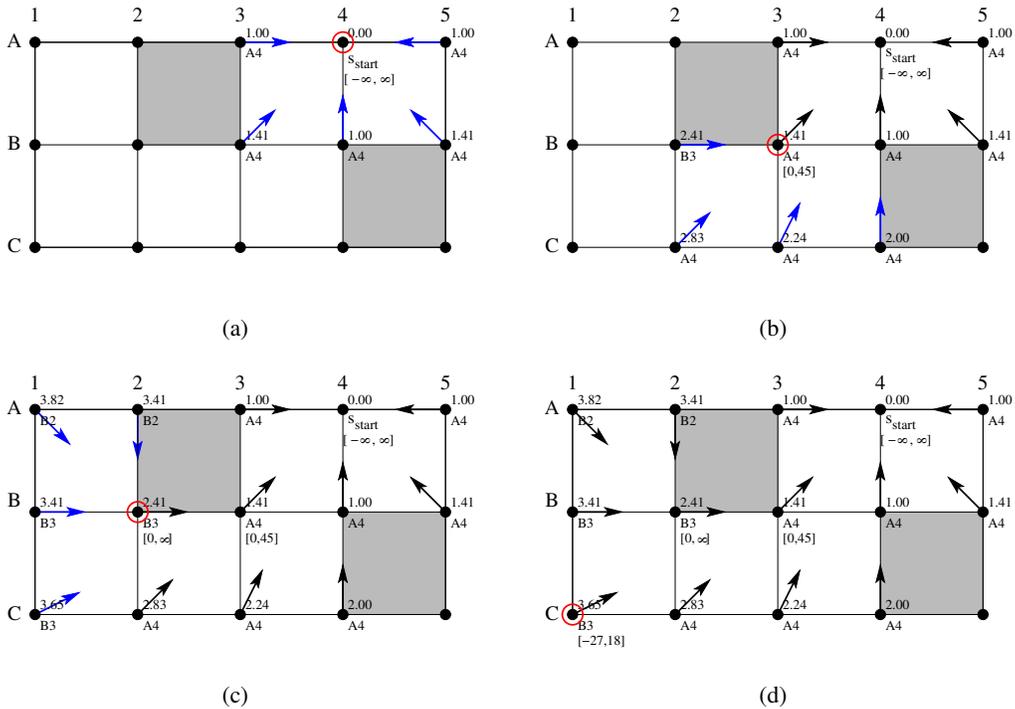

Figure 13: Example trace of AP Theta*

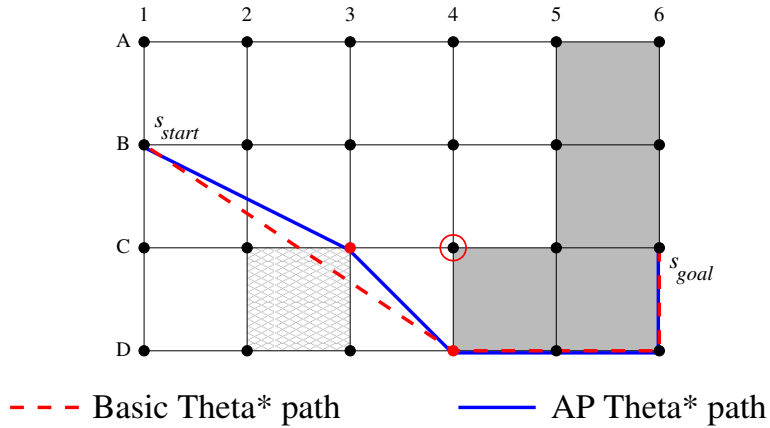

Figure 14: Basic Theta* path versus AP Theta* path

C3 because, for example, it does not know whether or not cell C2-C3-D3-D2 is unblocked. AP Theta* therefore cannot determine the angle range of vertex C4 exactly and makes the conservative assumption that vertex C3 barely has line-of-sight to vertex B1 and sets the lower angle bound of vertex C4 according to Case 4 based on vertex C3. It then uses the resulting angle range to determine that the unexpanded visible neighbor D4 of vertex C4 is not guaranteed to have line-of-sight to vertex B1. However, vertex D4 does have line-of-sight to vertex B1 if cell C2-C3-D3-D2 is unblocked. AP Theta* eventually finds the solid blue path [B1, C3, D4] from start vertex B1 to vertex D4, while Basic Theta* finds the dashed red path [B1, D4], which is the true shortest path.

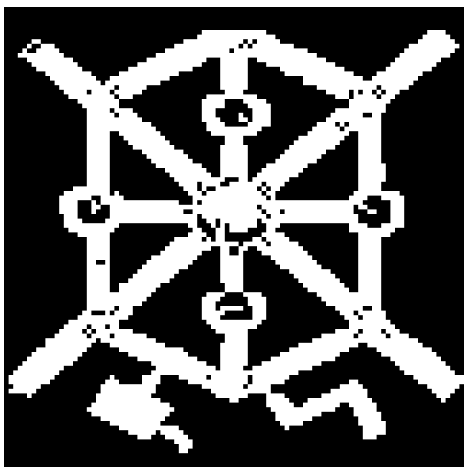

Figure 15: Map of Baldur's Gate II

The correctness and completeness proof of Basic Theta* needs to get changed slightly for AP Theta* since AP Theta* performs its line-of-sight checks differently.

**Theorem 2.** *AP Theta* terminates and path extraction retrieves an unblocked path from the start vertex to the goal vertex if such a path exists. Otherwise, AP Theta* terminates and reports that no unblocked path exists.*

*Proof.* The proof is similar to the proof of Theorem 1 since AP Theta* uses the angle ranges only to determine whether or not Path 2 is blocked but not to determine whether or not Path 1 is blocked. The only property that needs to be proved differently is that two vertices indeed have line-of-sight if (but not necessarily only if) the line-of-sight check of AP Theta* succeeds, see Appendix B. □

## 7. Experimental Results

In this section, we compare Basic Theta* and AP Theta* to A* on grids, A* PS, FD* and A* on visibility graphs with respect to their path length, number of vertex expansions, runtime (measured in seconds) and number of heading changes.

We compare these path-planning algorithms on $100 \times 100$ and $500 \times 500$ grids with different percentages of randomly blocked cells (random grids) and scaled maps from the real-time strategy game Baldur's Gate II (game maps). Figure 15 (Bulitko, Sturtevant, & Kazakevich, 2005) shows an example of a game map. The start and goal vertices are the south-west corners of cells. For random grids, the start vertex is in the south-west cell. The goal vertex is in a cell randomly chosen from the column of cells furthest east. Cells are blocked randomly but a one-unit border of unblocked cells guarantees that there is path from the start vertex to the goal vertex. For game maps, the start and goal vertices are randomly chosen from the corners of unblocked cells. We average over 500 random $100 \times 100$ grids, 500 random $500 \times 500$ grids and 118 game maps.

| | | FD* | Basic Theta* | AP Theta* | A* on Visibility Graphs (true shortest path) | A* on Grids | A* PS |
|---|---|---|---|---|---|---|---|
| 100×100 | Game Maps | 40.04 | 39.98 | 40.05 | 39.96 | 41.77 | 40.02 |
| | Random Grids 0% | 114.49 | 114.33 | 114.33 | 114.33 | 120.31 | 114.33 |
| | Random Grids 5% | 114.15 | 113.94 | 113.94 | 113.83 | 119.76 | 114.71 |
| | Random Grids 10% | 114.74 | 114.51 | 114.51 | 114.32 | 119.99 | 115.46 |
| | Random Grids 20% | 115.20 | 114.93 | 114.95 | 114.69 | 120.31 | 116.16 |
| | Random Grids 30% | 115.45 | 115.22 | 115.25 | 114.96 | 120.41 | 116.69 |
| 500×500 | Game Maps | 223.64 | 223.30 | 224.40 | N/A | 233.66 | 223.70 |
| | Random Grids 0% | 576.19 | 575.41 | 575.41 | N/A | 604.80 | 575.41 |
| | Random Grids 5% | 568.63 | 567.30 | 567.34 | N/A | 596.45 | 573.46 |
| | Random Grids 10% | 576.23 | 574.57 | 574.63 | N/A | 603.51 | 581.03 |
| | Random Grids 20% | 580.19 | 578.41 | 578.51 | N/A | 604.93 | 585.62 |
| | Random Grids 30% | 581.73 | 580.18 | 580.35 | N/A | 606.38 | 588.98 |

Table 1: Path length

| | | FD* | Basic Theta* | AP Theta* | A* on Visibility Graphs (true shortest path) | A* on Grids | A* PS |
|---|---|---|---|---|---|---|---|
| 100×100 | Game Maps | 0.0111 | 0.0060 | 0.0084 | 0.4792 | 0.0048 | 0.0052 |
| | Random Grids 0% | 0.0229 | 0.0073 | 0.0068 | 0.0061 | 0.0053 | 0.0208 |
| | Random Grids 5% | 0.0275 | 0.0090 | 0.0111 | 0.0766 | 0.0040 | 0.0206 |
| | Random Grids 10% | 0.0305 | 0.0111 | 0.0145 | 0.3427 | 0.0048 | 0.0204 |
| | Random Grids 20% | 0.0367 | 0.0150 | 0.0208 | 1.7136 | 0.0084 | 0.0222 |
| | Random Grids 30% | 0.0429 | 0.0183 | 0.0263 | 3.7622 | 0.0119 | 0.0240 |
| 500×500 | Game Maps | 0.1925 | 0.1166 | 0.1628 | N/A | 0.0767 | 0.1252 |
| | Random Grids 0% | 0.3628 | 0.1000 | 0.0234 | N/A | 0.0122 | 0.6270 |
| | Random Grids 5% | 0.4514 | 0.1680 | 0.1962 | N/A | 0.0176 | 0.6394 |
| | Random Grids 10% | 0.5608 | 0.2669 | 0.3334 | N/A | 0.0573 | 0.6717 |
| | Random Grids 20% | 0.6992 | 0.3724 | 0.5350 | N/A | 0.1543 | 0.6852 |
| | Random Grids 30% | 0.8562 | 0.5079 | 0.7291 | N/A | 0.3238 | 0.7355 |

Table 2: Runtime

| | | FD* | Basic Theta* | AP Theta* | A* on Visibility Graphs (true shortest path) | A* on Grids | A* PS |
|---|---|---|---|---|---|---|---|
| 100×100 | Game Maps | 247.07 | 228.45 | 226.42 | 68.23 | 197.19 | 315.08 |
| | Random Grids 0% | 592.74 | 240.42 | 139.53 | 1.00 | 99.00 | 1997.29 |
| | Random Grids 5% | 760.17 | 430.06 | 361.17 | 35.35 | 111.96 | 1974.27 |
| | Random Grids 10% | 880.21 | 591.31 | 520.91 | 106.23 | 169.98 | 1936.56 |
| | Random Grids 20% | 1175.42 | 851.79 | 813.14 | 357.33 | 386.41 | 2040.10 |
| | Random Grids 30% | 1443.44 | 1113.40 | 1089.96 | 659.36 | 620.18 | 2153.28 |
| 500×500 | Game Maps | 6846.62 | 6176.37 | 6220.58 | N/A | 5580.32 | 9673.88 |
| | Random Grids 0% | 11468.11 | 2603.40 | 663.34 | N/A | 499.00 | 49686.47 |
| | Random Grids 5% | 15804.81 | 7450.85 | 5917.25 | N/A | 755.66 | 49355.41 |
| | Random Grids 10% | 19874.62 | 11886.95 | 10405.34 | N/A | 2203.83 | 50924.01 |
| | Random Grids 20% | 26640.83 | 18621.61 | 17698.75 | N/A | 6777.15 | 50358.66 |
| | Random Grids 30% | 34313.28 | 25744.57 | 25224.92 | N/A | 14641.36 | 53732.82 |

Table 3: Number of vertex expansions

All path-planning algorithms are implemented in C# and executed on a 3.7 GHz Core 2 Duo with 2 GByte of RAM. Our implementations are not optimized and can possibly be improved.

| | | FD* | Basic Theta* | AP Theta* | A* on Visibility Graphs (true shortest paths) | A* on Grids | A* PS |
|---|---|---|---|---|---|---|---|
| 100×100 | Game Maps | 34.25 | 3.08 | 3.64 | 2.92 | 5.21 | 2.83 |
| | Random Grids 0% | 123.40 | 0.00 | 0.00 | 0.00 | 0.99 | 0.00 |
| | Random Grids 5% | 113.14 | 5.14 | 6.03 | 5.06 | 6.00 | 4.53 |
| | Random Grids 10% | 106.66 | 8.96 | 9.87 | 8.84 | 10.85 | 8.48 |
| | Random Grids 20% | 98.76 | 15.21 | 15.96 | 14.74 | 19.42 | 14.45 |
| | Random Grids 30% | 96.27 | 19.96 | 20.62 | 19.44 | 26.06 | 18.35 |
| 500×500 | Game Maps | 219.70 | 4.18 | 7.58 | N/A | 10.19 | 3.84 |
| | Random Grids 0% | 667.00 | 0.00 | 0.00 | N/A | 1.00 | 0.00 |
| | Random Grids 5% | 592.65 | 21.91 | 27.99 | N/A | 24.68 | 22.27 |
| | Random Grids 10% | 559.69 | 41.60 | 47.40 | N/A | 49.73 | 43.16 |
| | Random Grids 20% | 506.10 | 72.49 | 76.79 | N/A | 91.40 | 69.44 |
| | Random Grids 30% | 481.16 | 97.21 | 100.31 | N/A | 123.81 | 89.43 |

Table 4: Number of heading changes

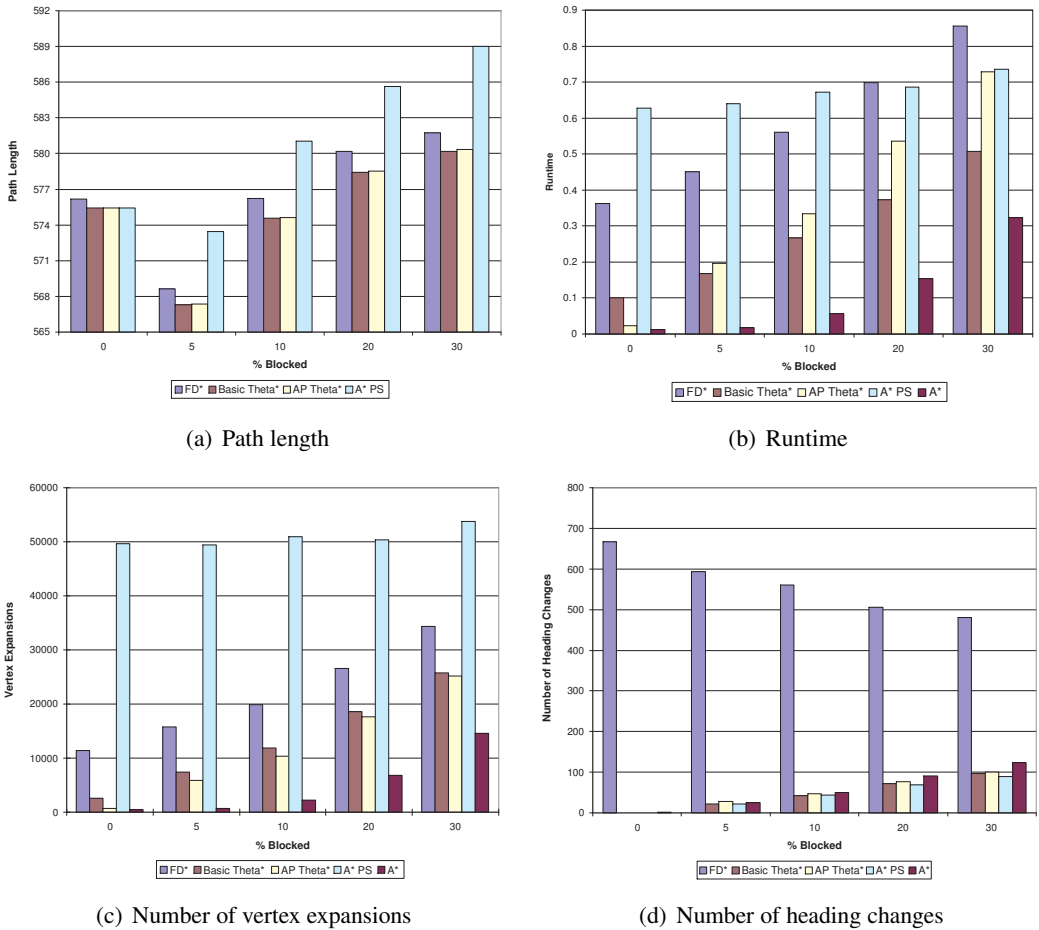

(a) Path length

(b) Runtime

(c) Number of vertex expansions

(d) Number of heading changes

Figure 16: Random 500 × 500 grids

```
 99  h(s)
100      Δ_x := |s.x − (s_goal).x|;
101      Δ_y := |s.y − (s_goal).y|;
102      largest := max(Δ_x, Δ_y);
103      smallest := min(Δ_x, Δ_y);
104      return √2 · smallest + (largest − smallest);
105  end
```

**Algorithm 5**: Calculation of octile distances

A* on grids, A* PS, FD* and A* on visibility graphs break ties among vertices with the same f-value in the open list in favor of vertices with larger g-values (when they decide which vertex to expand next) since this tie-breaking scheme typically results in fewer vertex expansions and thus shorter runtimes for A*. Care must thus be taken when calculating the g-values, h-values and f-values precisely. The numerical precision of these floating point numbers can be improved for A* on grids by representing them in the form $m + \sqrt{2}n$ for integers $m$ and $n$. Basic Theta* and AP Theta* break ties in favor of vertices with smaller g-values for the reasons explained in Section 9.

We use all path-planning algorithms with consistent h-values since consistent h-values result in short paths for A*. Consistent h-values satisfy the triangle inequality, that is, the h-value of the goal vertex is zero and the h-value of any potential non-goal parent of any vertex is no greater than the distance from the potential non-goal parent of the vertex to the vertex plus the h-value of the vertex (Hart et al., 1968; Pearl, 1984). Consistent h-values are lower bounds on the corresponding goal distances of vertices. Increasing consistent h-values typically decreases the number of vertex expansions for A* and thus also the runtime of A*. We thus use all path-planning algorithms with the largest consistent h-values that are easy to calculate. For Basic Theta*, AP Theta*, FD* and A* on visibility graphs, the goal distances of vertices can be equal to the true goal distances, that is, the goal distances on grids if the paths are not constrained to grid edges. We therefore use these path planning algorithms with the straight-line distances $h(s) = c(s, s_{goal})$ as h-values in our experiments. The straight-line distances are the goal distances on grids without blocked cells if the paths are not constrained to grid edges. For A* on grids and A* PS, the goal distances of vertices are equal to the goal distances on grids if the paths are constrained to grid edges. We could therefore use them with the larger octile distances as h-values in our experiments. The octile distances are the goal distances on grids without blocked cells if the paths are constrained to grid edges. Algorithm 5 shows how to calculate the octile distance of a given vertex $s$, where $s.x$ and $s.y$ are the x and y coordinates of vertex $s$, respectively. We indeed use A* on grids with the octile distances but A* PS with the straight-line distances since smoothing is then typically able to shorten the resulting paths much more at an increase in the number of vertex expansions and thus runtime. Grids without blocked cells provide an example. With the octile distances as h-values, A* on grids finds paths in which all diagonal movements (whose lengths are $\sqrt{2}$) precede all horizontal or vertical movements (whose lengths are 1) because the paths with the largest number of diagonal movements are the longest ones among all paths with the same number of movements due to the tie-breaking scheme used. On the other hand, with the straight-line distances as h-values, A* on grids finds paths that interleave the diagonal movements with the horizontal and vertical movements (which means that it is likely that there are lots of opportunities to smooth the paths even for grids with some blocked cells) and that are closer to the straight line between the start and goal vertices (which means that it is likely that the paths are closer to true shortest paths even for grids with some blocked cells), because the h-values of vertices closer to the straight line are typically smaller than the h-values of vertices farther away from the straight line.

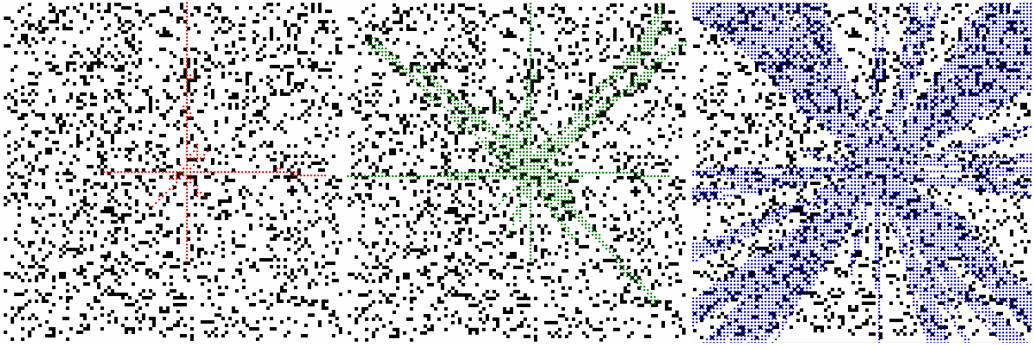

Figure 17: True shortest paths found by FD* (left), A* PS (middle) and Basic Theta* (right)

Tables 1-4 report our experimental results. The runtime of A* on visibility graphs (which finds true shortest paths) is too long on $500 \times 500$ grids and thus is omitted. Figure 16 visualizes the experimental results on random $500 \times 500$ grids. The path length of A* on grids is much larger than the path lengths of the other path-planning algorithms and thus is omitted.

We make the following observations about the path lengths:

- The path-planning algorithms in order of increasing path lengths tend to be: A* on visibility graphs (which finds true shortest paths), Basic Theta*, AP Theta*, FD*, A* PS and A* on grids. On random $500 \times 500$ grids with 20 percent blocked cells, Basic Theta* finds shorter paths than AP Theta* 70 percent of the time, shorter paths than FD* 97 percent of the time, shorter paths than A* PS 94 percent of the time and shorter paths than A* on grids 99 percent of the time.

- The paths found by Basic Theta* and AP Theta* are almost as short as true shortest paths even though AP Theta* sometimes constrains the angle ranges more than necessary. For example, they are on average less than a factor of 1.003 longer than true shortest paths on $100 \times 100$ grids.

- Basic Theta* finds true shortest paths more often than FD* and A* PS. Figure 17 shows an example where the light green vertex in the center is the start vertex and the red, green and blue vertices represent goal vertices to which FD*, A* PS and Basic Theta* find true shortest paths, respectively.

We make the following observations about the runtimes. The path-planning algorithms in order of increasing runtimes tend to be: A* on grids, Basic Theta*, AP Theta*, A* PS, FD* and A* on visibility graphs.

We make the following observations about the numbers of vertex expansions. The path-planning algorithms in order of increasing numbers of vertex expansions tend to be: A* on visibility graphs, A* on grids, AP Theta*, Basic Theta*, FD* and A* PS. (The number of vertex expansions of A* on grids and A* PS are different because we use them with different h-values.)

|  | FD* | Basic Theta* | AP Theta* | A* PS |
|---|---|---|---|---|
| Runtime | 5.21 | 3.65 | 5.70 | 3.06 |
| Runtime per Vertex Expansion | 0.000021 | 0.000015 | 0.000023 | 0.000012 |

Table 5: Path-planning algorithms without post-processing steps on random $500 \times 500$ grids with 20 percent blocked cells

Finally, we make the following observations about the number of heading changes. The path-planning algorithms in order of increasing numbers of heading changes tend to be: A* PS, A* on visibility graphs, Basic Theta*, AP Theta*, A* on grids and FD*.

There are some exceptions to the trends reported above. We therefore perform paired t-tests. They show with confidence level $\alpha = 0.01$ that Basic Theta* indeed finds shorter paths than AP Theta*, A* PS and FD* and that Basic Theta* indeed has a shorter runtime than AP Theta*, A* PS and FD*.

To summarize, A* on visibility graphs finds true shortest paths but is slow. On the other hand, A* on grids finds long paths but is fast. Any-angle path planning lies between these two extremes. Basic Theta* dominates AP Theta*, A* PS and FD* in terms of the tradeoff between runtime and path length. It finds paths that are almost as short as true shortest paths and is almost as fast as A* on grids. It is also simpler to implement than AP Theta*. Therefore, we build on Basic Theta* for the remainder of this article, although we report some experimental results for AP Theta* as well. However, AP Theta* reduces the runtime of Basic Theta* per vertex expansion from linear to constant. It is currently unknown whether or not constant time line-of-sight checks can be devised that make AP Theta* faster than Basic Theta*. This is an interesting area of future research since AP Theta* is potentially a first step toward significantly reducing the runtime of any-angle path planning via more sophisticated line-of-sight checks.

## 8. Extensions of Theta*

In this section, we extend Basic Theta* to find paths from a given start vertex to all other vertices and to find paths on grids that contain unblocked cells with non-uniform traversal costs.

### 8.1 Single Source Paths

So far, Basic Theta* has found paths from a given start vertex to a given goal vertex. We now discuss a version of Basic Theta* that finds single source paths (that is, paths from a given start vertex to all other vertices) by terminating only when the open list is empty instead of when either the open list is empty or it expands the goal vertex.

Finding single source paths requires all path-planning algorithms to expand the same number of vertices, which minimizes the influence of the h-values on the runtime and thus results in a clean comparison since the h-values sometimes are chosen to trade off between runtime and path length.

The runtimes of A* PS and FD* are effected more than those of Basic Theta* and AP Theta* when finding single source paths since they require post-smoothing or path-extraction steps for each

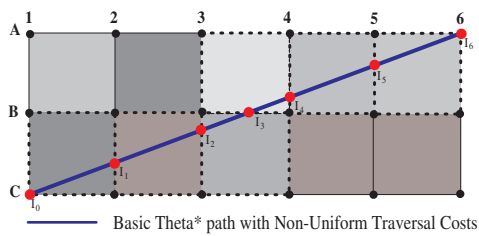

Figure 18: Basic Theta* on grids that contain unblocked cells with non-uniform traversal costs

(a) Small contiguous regions of uniform traversal costs

|  | A* on Grids | FD* | Basic Theta* |
|---|---|---|---|
| Path Cost | 4773.59 | 4719.26 | 4730.96 |
| Runtime | 11.28 | 14.98 | 19.02 |

(b) Large contiguous regions of uniform traversal costs

|  | A* on Grids | FD* | Basic Theta* |
|---|---|---|---|
| Path Cost | 1251.88 | 1208.89 | 1207.06 |
| Runtime | 3.42 | 5.31 | 5.90 |

Table 6: Path-planning algorithms on random $1000 \times 1000$ grids with non-uniform traversal costs

path, and thus need to post-process many paths. Table 5 reports the runtimes of the path-planning algorithms without these post-processing steps. The runtime of Basic Theta* per vertex expansion is similar to that of A* PS and shorter than that of either AP Theta* and FD* because the later two algorithms require more floating point operations.

## 8.2 Non-Uniform Traversal Costs

So far, Basic Theta* has found paths on grids that contain unblocked cells with uniform traversal costs. In this case, true shortest paths have heading changes only at the corners of blocked cells and the triangle inequality holds, which means that Path 2 is no longer than Path 1. We now discuss a version of Basic Theta* that finds paths on grids that contain unblocked cells with non-uniform traversal costs by computing and comparing path lengths (which are now path costs) appropriately. In this case, true shortest paths can also have heading changes at the boundaries between unblocked cells with different traversal costs and the triangle inequality is no longer guaranteed to hold, which means that Path 2 can be more costly than Path 1. Thus, Basic Theta* no longer unconditionally chooses Path 2 over Path 1 if Path 2 is unblocked [Line 42] but chooses the path with the smaller cost. It uses the standard Cohen-Sutherland clipping algorithm from computer graphics (Foley, van Dam, Feiner, & Hughes, 1992) to calculate the cost of Path 2 during the line-of-sight check. Figure 18 shows an example for the path segment $\overline{C1A6}$ from vertex C1 to vertex A6. This straight line is split into line segments at the points where it intersects with cell boundaries. The cost of the path segment is the sum of the costs of its line segments $\overline{I_i I_{i+1}}$, and the cost of each line segment is the product of its length and the traversal cost of the corresponding unblocked cell.

We found that changing the test on Line 52 in Algorithm 3 from "strictly less than" to "less than or equal to" slightly reduces the runtime of Basic Theta*. This is a result of the fact that it is faster to compute the cost of a path segment that corresponds to Path 1 than Path 2 since it tends to consist of fewer line segments.

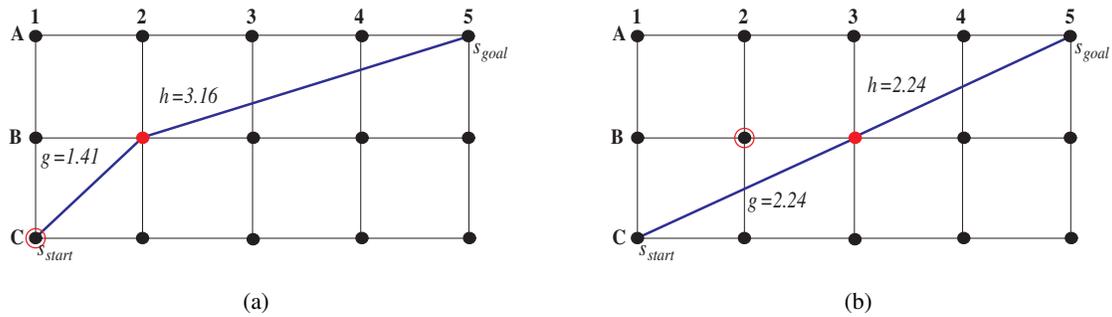

Figure 19: Non-monotonicity of f-values of Basic Theta*

We compare Basic Theta* to A* on grids and FD* with respect to their path cost and runtime (measured in seconds) since A* can easily be adapted to grids that contain unblocked cells with non-uniform traversal costs and FD* was designed for this case. We compare these path-planning algorithms on $1000 \times 1000$ grids, where each cell is assigned an integer traversal cost from 1 to 15 (corresponding to an unblocked cell) and infinity (corresponding to a blocked cell), similar to the technique used in the work of Ferguson and Stentz (2006) . If a path lies on the boundary between two cells with different traversal costs, then we use the smaller traversal cost of the two cells. The start and goal vertices are the south-west corners of cells. The start vertex is in the south-west cell. The goal vertex is in a cell randomly chosen from the column of cells furthest east. We average over 100 random grids. Table 6 (a) reports our results if every traversal cost is chosen with uniform probability, resulting in small contiguous regions of uniform traversal costs. The path cost and runtime of FD* are both smaller than those of Basic Theta*. The path cost of A* on grids is only about 1 percent larger than that of FD* although its runtime is much smaller than that of FD*. Thus, any-angle planning does not have a large advantage over A* on grids. Table 6(b) reports our results if traversal cost one is chosen with probability 50 percent and all other traversal costs are chosen with uniform probability, resulting in large contiguous regions of uniform traversal costs. The path cost of Basic Theta* is now smaller than that of FD* and its runtime is about the same as that of FD*. The paths found by FD* tend to have many more unnecessary heading changes in regions with the same traversal costs than those of Basic Theta*, which outweighs the paths found by Basic Theta* not having necessary heading changes on the boundary between two cells with different traversal costs. The path cost of A* on grids is more than 3 percent larger than that of Basic Theta*. Thus, any-angle planning now has a larger advantage over A* on grids.

## 9. Trading Off Runtime and Path Length: Exploiting h-Values

There are strategies for trading off runtime and path length that A* on grids and Basic Theta* share. However, their behavior can be very different even though the two algorithms have very similar pseudocode. In this section, we develop versions of Basic Theta* that might be able to find shorter paths at an increase in runtime, including versions that use weighted h-values with weights less than one, that break ties among vertices with the same f-value in the open list in favor of vertices with smaller g-values (when they decide which vertex to expand next) and that re-expand vertices whose f-values have decreased.

We use all path-planning algorithms with consistent h-values. A* on grids then has the following properties (Pearl, 1984): The f-value of any expanded vertex is no larger than the f-value of any of its unexpanded visible neighbors after updating them according to Path 1, which implies that the f-value of any vertex that is expanded before some other vertex is no larger than the f-value of this other vertex. Consequently, at any point in time during a search once a vertex has been expanded, following the parents from the expanded vertex to the start vertex retrieves a shortest path from the start vertex to the expanded vertex in reverse, which implies that A* cannot find shorter paths by expanding vertices more than once. Basic Theta* has different properties: The f-value of an expanded vertex can be larger than the f-value of one or more of its unexpanded visible neighbors after updating them according to Path 2, which implies that the f-value of a vertex that is expanded before some other vertex can be larger than the f-value of this other vertex. Consequently, at any point in time during a search once a vertex has been expanded, following the parents from the expanded vertex to the start vertex is not guaranteed to retrieve a shortest path from the start vertex to the vertex in reverse, which implies that Basic Theta* might find shorter paths by expanding vertices more than once. Figure 19 shows an example. When Basic Theta* expands start vertex C1 with parent C1, it generates vertex B2. Vertex B2 is an unexpanded visible neighbor of vertex C1 that has line-of-sight to vertex C1. Basic Theta* thus updates it according to Path 2 (which is the same as Path 1 in this case), sets its f-value to $f(B2) = 1.41 + 3.16 = 4.57$, sets its parent to vertex C1 and inserts it into the open list (Figure 19(a)). When Basic Theta* later expands vertex B2 with parent C1, it generates vertex B3. Vertex B3 is an unexpanded visible neighbor of vertex B2 that has line-of-sight to vertex C1. Basic Theta* thus updates it according to Path 2, sets its f-value to $f(B3) = 2.24 + 2.24 = 4.48$, sets its parent to vertex C1 and inserts it into the open list (Figure 19(b)). Thus, the f-value of expanded vertex B2 is indeed larger than the f-value of its unexpanded visible neighbor B3 after updating it according to Path 2 because the increase in g-value from vertex B2 to vertex B3 [= 0.83] is less than the decrease in h-value from vertex B2 to vertex B3 [= 0.92]. When Basic Theta* later expands vertex B3, the f-value of vertex B2 [= 4.57] that is expanded before vertex B3 is indeed larger than the f-value of vertex B3 [= 4.48].

These properties suggest that Basic Theta* might be able to find shorter paths at an increase in runtime by re-expanding vertices or expanding additional vertices (for example by using weighted h-values with weights less than one) while A* cannot. At the same time, standard optimizations of A* that decrease its runtime might also be able to decrease the runtime of Basic Theta* (such as breaking ties among vertices with the same f-value in the open list in favor of vertices with larger g-values). In this section we investigate these tradeoffs.

## 9.1 Weighted h-Values

So far, Basic Theta* has used consistent h-values $h(s)$. A* with consistent h-values finds paths of the same length no matter how small or large the h-values are. Decreasing consistent h-values typically increases the number of vertex expansions for A*. We therefore now discuss a version of Basic Theta* that might be able to find shorter paths at an increase in runtime by using weighted h-values with weights less than one. This version of Basic Theta* uses the h-values $h(s) = w \times c(s, s_{goal})$ for a given weight $0 \leq w < 1$ and thus is similar to Weighted A* (Pohl, 1973), except that Weighted A* typically uses weights greater than one. Figure 20(a) shows an example of the resulting effect on the number of vertex expansions and path length. The green vertex in the north-east is the start

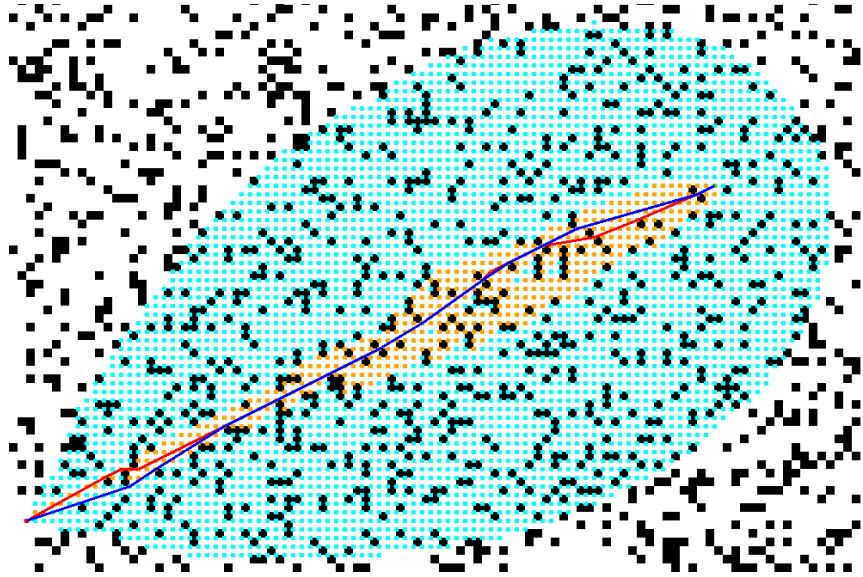

(a) Expanded vertices by Basic Theta* with different weights

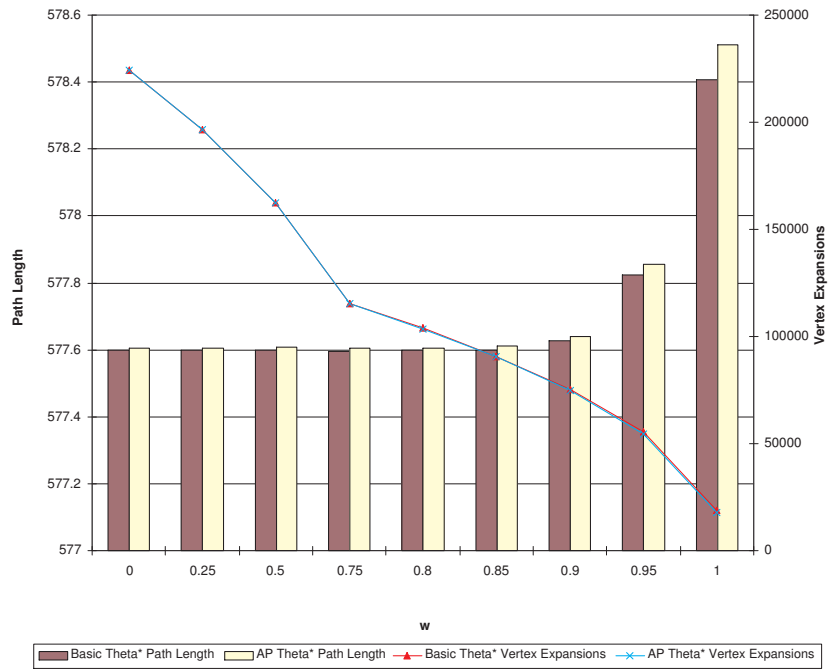

(b) Random $500 \times 500$ grids with 20 percent blocked cells

Figure 20: Weighted h-values

vertex, and the red vertex in the south-west is the goal vertex. Basic Theta* with weight 1.00 (as used so far) expands the orange vertices and finds the red path. Basic Theta* with weight 0.75 expands the blue vertices and finds the blue path. Thus, Basic Theta* expands more vertices with weight 0.75 than with weight 1.00 and the resulting path is shorter since it passes through vertices that are expanded with weight 0.75 but not with weight 1.00.

|  | Smaller g-Values | | Larger g-Values | |
|---|---|---|---|---|
|  | Basic Theta* | AP Theta* | Basic Theta* | AP Theta* |
| Path Length | 578.41 | 578.51 | 578.44 | 578.55 |
| Number of Vertex Expansions | 18621.61 | 17698.75 | 18668.03 | 17744.94 |
| Runtime | 0.3724 | 0.5350 | 0.3829 | 0.5389 |

Table 7: Random $500 \times 500$ grids with 20 percent blocked cells

Figure 20(b) reports the effect of different weights on the path length and number of vertex expansions of Basic Theta* and AP Theta* on random $500 \times 500$ grids with 20 percent blocked cells. (The graphs of the number of vertex expansions of Basic Theta* and AP Theta* nearly coincide.) Decreasing the weight decreases the path length at an increase in the number of vertex expansions and thus the runtime. The path length decreases more for AP Theta* than Basic Theta* since AP Theta* can constrain the angle ranges more than necessary and thus benefits in two ways from expanding more vertices. However, neither Basic Theta* nor AP Theta* are guaranteed to find true shortest paths even if their weights are zero.

## 9.2 Tie Breaking

So far, Basic Theta* has broken ties among vertices in the open list with the same f-value in favor of vertices with larger g-values (when it decides which vertex to expand next). A* with consistent h-values finds paths of the same length no matter which tie-breaking scheme it uses. Breaking ties in favor of vertices with smaller g-values typically increases the number of vertex expansions and thus the runtime. We therefore discuss a version of Basic Theta* that might be able to find shorter paths at an increase in runtime by breaking ties in favor of vertices with smaller g-values. Figure 21 shows an example of the resulting effect on path length. Vertices C4 and B4 have the same f-value but vertex B4 has a larger g-value since $f(C4) = 3.83 + 1.41 = 5.24$ and $f(B4) = 4.24 + 1 = 5.24$. If Basic Theta* breaks ties in favor of vertices with larger g-values, then it expands vertex B4 with parent E1 before vertex C4 with parent C3 and eventually expands the goal vertex with parent B4 and terminates. Path extraction then follows the parents from goal vertex B5 to start vertex E1 to retrieve the dashed red path [E1, B4, B5]. However, if Basic Theta* breaks ties in favor of vertices with smaller g-values, then it expands vertex C4 with parent C3 before vertex B4 with parent E1 and eventually expands the goal vertex with parent C3 and terminates. Path extraction then follows the parents from goal vertex B5 to start vertex E1 to retrieve the shorter solid blue path [E1, C3, B5].

Table 7 reports the effect of the tie-breaking scheme on the path length, number of vertex expansions and runtime of Basic Theta* and AP Theta* on random $500 \times 500$ grids with 20 percent blocked cells. Breaking ties in favor of vertices with smaller g-values neither changes the path length, number of vertex expansions nor runtime significantly. The effect of the tie-breaking scheme is small since fewer vertices have the same f-value for Basic Theta* and AP Theta* than for A* on grids because the number of possible g-values and h-values is larger for any-angle path planning.

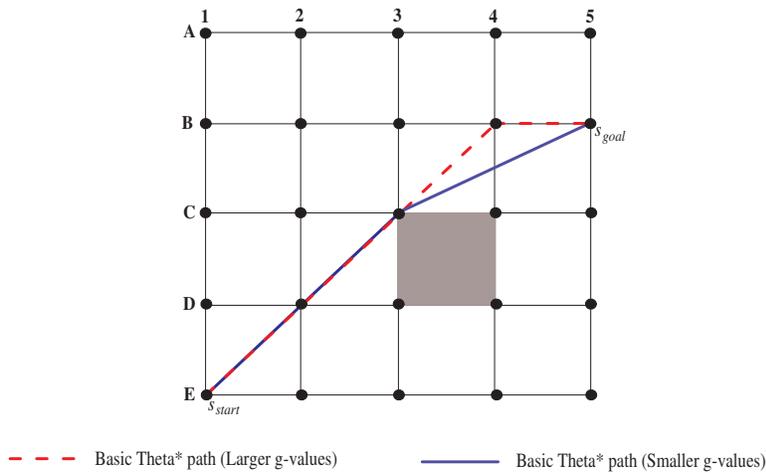

Figure 21: Basic Theta* paths for different tie-breaking schemes

|  | Basic Theta* without Vertex Re-Expansions | Basic Theta* with Vertex Re-Expansions |
|---|---|---|
| Path Length | 578.41 | 577.60 |
| Number of Vertex Expansions | 18621.61 | 22836.37 |
| Runtime | 0.3724 | 0.5519 |

Table 8: Random $500 \times 500$ grids with 20 percent blocked cells

There is also a second method in which breaking ties can effect path length. So far, Basic Theta* has chosen Path 2 over Path 1 if an unexpanded visible neighbor of a vertex has line-of-sight to the parent of the vertex. However, it can choose Path 1 over Path 2 if both paths are equally long, which increases the runtime due to the additional comparison. Figure 21 shows an example of the resulting effect on path length. Assume that Basic Theta* expands vertex B4 before vertex C4. If Basic Theta* chooses Path 2 over Path 1 then it expands vertex B4 with parent E1 and eventually expands the goal vertex B5 with parent B4 and terminates. Path extraction then follows the parents from goal vertex B5 to start vertex E1 to retrieve the dashed red path [E1, B4, B5]. However, if Basic Theta* chooses Path 1 over Path 2 then it expands vertex B4 with parent C3 and eventually expands goal vertex B5 with parent C3 and terminates. Path extraction then follows the parents from goal vertex B5 to start vertex E1 to retrieve the shorter solid blue path [E1, C3, B5].

## 9.3 Re-Expanding Vertices

So far, Basic Theta* has used a closed list to ensure that it expands each vertex at most once. A* with consistent h-values does not re-expand vertices whether or not it uses a closed list since it cannot find a shorter path from the start vertex to a vertex after expanding that vertex. On the other hand, Basic Theta* can re-expand vertices if it does not use a closed list since it can find a shorter path from the start vertex to a vertex after expanding the vertex. It then re-inserts the vertex into the open list and eventually re-expands it.[2] Figure 22 shows an example of the effect of vertex re-expansions on path length. Basic Theta* without vertex re-expansions eventually expands vertex C8 with parent D4. Vertex C9 is an unexpanded visible neighbor of vertex C8 that has line-of-sight to vertex D4. Basic Theta* without vertex re-expansions thus updates it according to Path 2 and sets its parent to vertex D4. After termination, path extraction follows the parents from goal vertex B9 to start vertex E1 to retrieve the dashed red path [E1, D4, C9, B9]. However, Basic Theta* with vertex re-expansions eventually expands vertex C8 with parent D4 and later re-expands vertex C8 with parent E1. Vertex C9 is a visible neighbor of vertex C8 that has line-of-sight to vertex E1. Basic Theta* with vertex re-expansions thus updates it according to Path 2 and sets its parent to vertex E1. After termination, path extraction follows the parents from goal vertex B9 to start vertex E1 to retrieve the shorter solid blue path [E1, C9, B9].

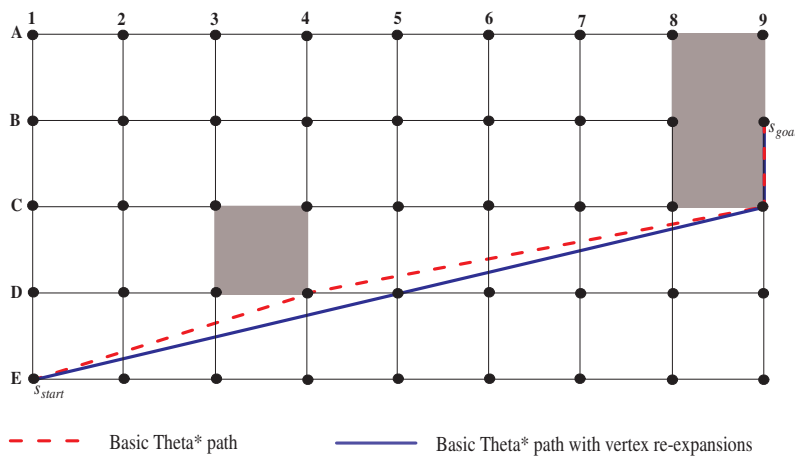

Figure 22: Basic Theta* paths with and without vertex re-expansions

**Theorem 3.** *Basic Theta\* with vertex re-expansions terminates and path extraction returns an unblocked path from the start vertex to the goal vertex if such a path exists. Otherwise, Basic Theta\* with vertex re-expansions terminates and reports that no unblocked path exists.*

*Proof.* The proof is similar to the proof of Theorem 1. The only property that needs to be proved differently is that Basic Theta* with vertex re-expansions terminates since it is no longer true that it can never insert a vertex into the open list again once it has removed the vertex from the open list. However, since the number of vertices is finite, there are only a finite number of acyclic paths from the start vertex to each vertex. Therefore, the number of possible g-values is finite. Therefore, Basic Theta* with vertex re-expansions can reduce the g-value of each vertex only a finite number of times and thus inserts each vertex into the open list a finite number of times. Thus, the open list eventually becomes empty and Basic Theta* has to terminate if it has not terminated earlier already. ☐

---

2. Basic Theta* with vertex re-expansions could also delay the expansion of the goal vertex (for example, by increasing its f-value artificially) so that it can re-expand more vertices before it terminates but our version of Basic Theta* with vertex re-expansions does not do that.

Table 8 reports the effect of vertex re-expansions on the path length, number of vertex expansions and runtime of Basic Theta* on random $500 \times 500$ grids with 20 percent blocked cells. Vertex re-expansions decrease the path length slightly at an increase in the number of vertex expansions and thus the runtime.

## 10. Trading Off Runtime and Path Length: Other Approaches

There are additional strategies for trading off runtime and path length that are specific to Basic Theta*. In this section, we develop versions of Basic Theta* that might be able to find shorter paths at an increase in runtime by examining more paths, including versions that check for line-of-sight to the parent of a parent, that use key vertices to identify promising parents and that increase the number of visible neighbors and thus the number of potential parents when updating vertices according to Path 1.

### 10.1 Three Paths

So far, Basic Theta* has considered two paths (namely Paths 1 and 2) when it updates the g-value and parent of an unexpanded visible neighbor $s'$ of vertex $s$. We now discuss a version of Basic Theta* that considers a third path, namely the path from the start vertex to the parent of the parent of vertex $s$ [$= g(parent(parent(s)))$] and from it to vertex $s'$ in a straight line [$= c(parent(parent(s)), s')$], resulting in a length of $g(parent(parent(s))) + c(parent(parent(s)), s')$. This version of Basic Theta* might be able to find shorter paths at an increase in runtime since the third path is no longer than Path 2 due to the triangle inequality. However, our experimental results (not reported here) show that the third path does not decrease the path length significantly because the original version of Basic Theta* already determines that the parent of the parent of vertex $s$ does not have line-of-sight to some vertex that shares its parent with vertex $s$. Thus, it is very unlikely that the parent of the parent of vertex $s$ has line-of-sight to vertex $s'$ and thus that the third path is unblocked.

### 10.2 Key Vertices

So far, Basic Theta* has considered two paths (namely Paths 1 and 2) when it updates the g-value and parent of an unexpanded visible neighbor $s'$ of vertex $s$. The parent of a vertex then is either a visible neighbor of the vertex or the parent of a visible neighbor, which is not always the case for true shortest paths. We now discuss a version of Basic Theta* that considers additional paths, namely the paths from the start vertex to cached key vertices and from them to vertex $s'$ in a straight line. This version of Basic Theta* might be able to find shorter paths at an increase in runtime due to the fact that the parent of a vertex can now also be one of the key vertices. However, our experimental results (not reported here) show that key vertices decrease the path length only slightly at a larger increase in runtime due to the overhead of having to select key vertices, maintain them and consider a larger number of paths.

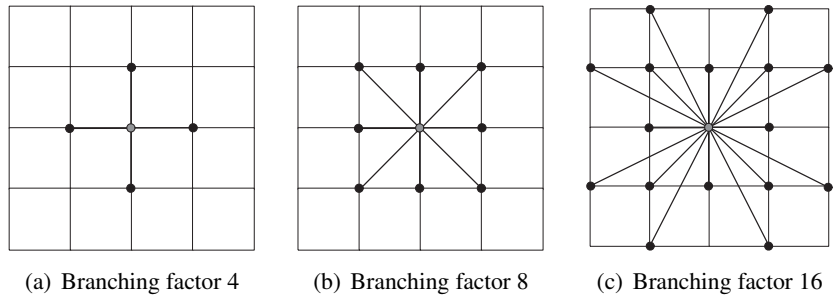

(a) Branching factor 4          (b) Branching factor 8          (c) Branching factor 16

Figure 23: Grids with different branching factors

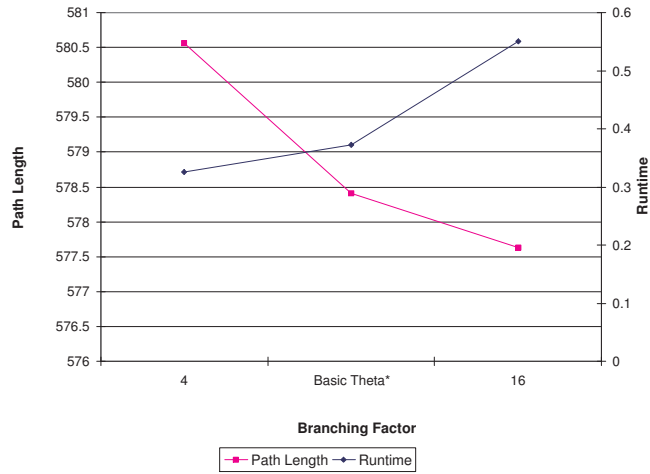

Figure 24: Basic Theta* on random $500 \times 500$ grids with 20 percent blocked cells

## 10.3 Larger Branching Factors

So far, Basic Theta* has operated on eight-neighbor grids. We now discuss a version of Basic Theta* that operates on grids with different numbers of neighbors and thus different branching factors. Figure 23 shows the neighbors of the center vertex for branching factors 4, 8 and 16 respectively. This version of Basic Theta* might be able to find shorter paths at an increase in runtime since larger branching factors increase the number of visible neighbors of vertices and thus their number of potential parents when updating them according to Path 1. Figure 24 reports the effect of larger branching factors on the path length and runtime of Basic Theta* on random $500 \times 500$ grids with 20 percent blocked cells. Larger branching factors indeed decrease the path length at an increase in runtime.

## 11. Conclusions

Any-angle path-planning algorithms find paths without artificially constraining the headings of the paths. We presented two new correct and complete any-angle path-planning algorithms. Basic Theta* and Angle-Propagation Theta* (AP Theta*) are both variants of A* that propagate information along grid edges (to achieve a short runtime) without constraining paths to grid edges (to find any-angle paths). Basic Theta* is simple to understand and implement, fast and finds short paths. However, it is not guaranteed to find true shortest paths. AP Theta* achieves a worst-case complexity per vertex expansion that is constant (like that of A* on grids) rather than linear in the number of cells (like that of Basic Theta*) by propagating angle ranges when it expands vertices. However, AP Theta* is more complex than Basic Theta*, is not as fast and finds slightly longer paths.

We proved the correctness and completeness of Basic Theta* and AP Theta* and then compared them against three existing any-angle path-planning algorithms, namely A* with post-smoothed paths (A* PS), A* on visibility graphs and Field D* (FD*), the only other version of A* we know of that propagates information along grid edges without constraining the paths to grid edges. Basic Theta* and AP Theta* (unlike A* PS) consider paths not constrained to grid edges during their search and thus can make informed decisions regarding these paths during the search. Basic Theta* and AP Theta* (unlike FD*) take advantage of the fact that true shortest paths have heading changes only at the corners of blocked cells.

A* on visibility graphs finds true shortest paths but is slow. On the other hand, A* on grids finds long paths but is fast. Any-angle path planning lies between these two extremes. Basic Theta* dominates AP Theta*, A* PS and FD* in terms of their tradeoffs between runtime and path length. It finds paths that are almost as short as true shortest paths and is almost as fast as A* on grids.

We extended Basic Theta* to find paths from a given start vertex to all other vertices and to find paths on grids that contain cells with non-uniform traversal costs. The f-value of an expanded vertex of Basic Theta* (unlike A* on grids) with consistent h-values can be larger than the f-value of one or more of its unexpanded visible neighbors, which means that Basic Theta* might be able to find shorter paths at an increase in runtime by re-expanding vertices or expanding additional vertices. We thus developed versions of Basic Theta* that use weighted h-values with weights less than one, that break ties among vertices with the same f-value in the open list in favor of vertices with smaller g-values (when they decide which vertex to expand next), that re-expand vertices whose f-values have decreased, that check for line-of-sight to the parent of a parent, that use key vertices to identify promising parents and that increase the number of visible neighbors.

In the future, we intend to develop a worst-case bound on the path lengths of Basic Theta* and AP Theta*, to better understand their properties and to investigate faster versions of AP Theta* that perform line-of-sight checks in constant time.

## Appendix A. Checking Line-of-Sight

In this appendix, we explain how to perform line-of-sight checks fast. For simplicity, we allow straight lines to pass between diagonally touching blocked cells. Performing a line-of-sight check is similar to determining which points to plot on a raster display when drawing a straight line between two points. The plotted points correspond to the cells that the straight line passes through.

```
106  LineOfSight(s, s')
107      x₀ := s.x;
108      y₀ := s.y;
109      x₁ := s'.x;
110      y₁ := s'.y;
111      d_y := y₁ − y₀;
112      d_x := x₁ − x₀;
113      f := 0;
114      if d_y < 0 then
115          d_y := −d_y;
116          s_y := −1;
117      else
118          s_y := 1;
119      if d_x < 0 then
120          d_x := −d_x;
121          s_x := −1;
122      else
123          s_x := 1;
124      if d_x ≥ d_y then
125          while x₀ ≠ x₁ do
126              f := f + d_y;
127              if f ≥ d_x then
128                  if grid(x₀ + ((s_x − 1)/2), y₀ + ((s_y − 1)/2)) then
129                      return false;
130                  y₀ := y₀ + s_y;
131                  f := f − d_x;
132              if f ≠ 0 AND grid(x₀ + ((s_x − 1)/2), y₀ + ((s_y − 1)/2)) then
133                  return false;
134              if d_y = 0 AND grid(x₀ + ((s_x − 1)/2), y₀) AND grid(x₀ + ((s_x − 1)/2), y₀ − 1) then
135                  return false;
136              x₀ := x₀ + s_x;
137      else
138          while y₀ ≠ y₁ do
139              f := f + d_x;
140              if f ≥ d_y then
141                  if grid(x₀ + ((s_x − 1)/2), y₀ + ((s_y − 1)/2)) then
142                      return false;
143                  x₀ := x₀ + s_x;
144                  f := f − d_y;
145              if f ≠ 0 AND grid(x₀ + ((s_x − 1)/2), y₀ + ((s_y − 1)/2)) then
146                  return false;
147              if d_x = 0 AND grid(x₀, y₀ + ((s_y − 1)/2)) AND grid(x₀ − 1, y₀ + ((s_y − 1)/2)) then
148                  return false;
149              y₀ := y₀ + s_y;
150      return true;
151  end
```

**Algorithm 6**: Line-of-sight algorithm

Thus, two vertices have line-of-sight iff none of the plotted points correspond to blocked cells. This allows Basic Theta* to perform its line-of-sight checks with the standard Bresenham line-drawing algorithm from computer graphics (Bresenham, 1965), that uses only fast logical and integer operations rather than floating-point operations. Algorithm 6 shows the resulting line-of-sight algorithm, where $s.x$ and $s.y$ are the x and y coordinates of vertex $s$, respectively, $grid$ represents the grid and $grid(x, y)$ is true iff the corresponding cell is blocked.

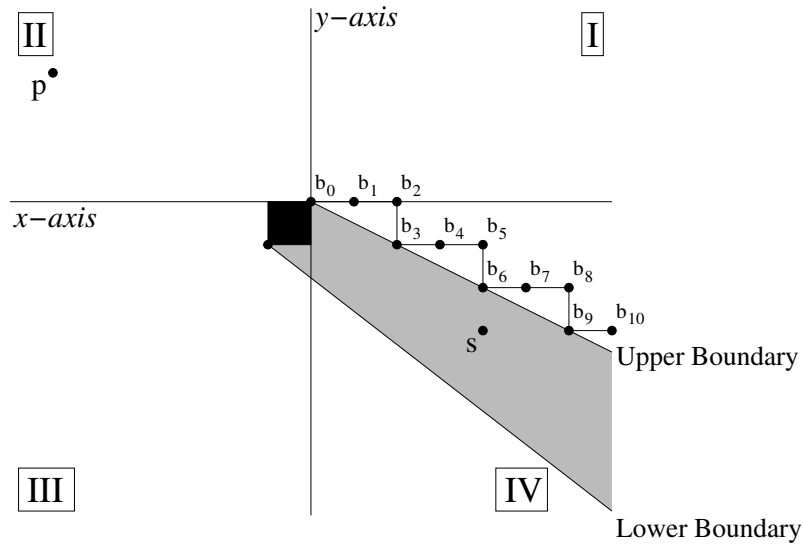

Figure 25: Parent, blocked cell and boundary vertices

## Appendix B. AP Theta* Returns Unblocked Paths

In this appendix, we prove that AP Theta* never returns a blocked path.

**Theorem 4.** *AP Theta* never returns a blocked path.*

*Proof.* We define a path to be blocked iff at least one vertex on the path does not have line-of-sight to its successor on the path. Thus, a path is blocked iff at least one of its path segments passes through the interior of a blocked cell or passes between two blocked cells that share an edge.

We first prove that AP Theta* never returns a path with a path segment that passes through the interior of a blocked cell. We prove by contradiction that AP Theta* cannot assign some parent $p$ to some vertex $s$ such that the path segment from parent $p$ to vertex $s$ passes through the interior of some blocked cell $b$. Assume otherwise. To simplify the proof, we translate and rotate the grid such that blocked cell $b$ is immediately south-west of the origin $b_0$ of the grid and parent $p$ is in quadrant II, as shown in Figure 25. We define the quadrant of a vertex $s$ as follows, where $s.x$ and $s.y$ are the x and y coordinates of vertex $s$, respectively:

- Quadrant I is the north-east quadrant (excluding the x-axis) given by $s.x \geq 0$ and $s.y > 0$.

- Quadrant II is the north-west quadrant (excluding the y-axis) given by $s.x < 0$ and $s.y \geq 0$.

- Quadrant III is the south-west quadrant (excluding the x-axis) given by $s.x \leq 0$ and $s.y < 0$.

572

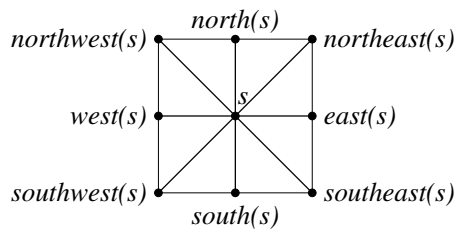

Figure 26: Neighbors of vertex $s$

- Quadrant IV is the south-east quadrant (excluding the y-axis but including the origin $b_0$) given by $s.x > 0$ and $s.y \leq 0$ or $s.x = 0$ and $s.y = 0$.

We refer to the neighbors of vertex $s$ as $east(s)$, $northeast(s)$, $north(s)$, $northwest(s)$, $west(s)$, $southwest(s)$, $south(s)$, $southeast(s)$, as shown in Figure 26.

Assume that there is a light source at vertex $p$ and that light cannot pass through blocked cell $b$, which creates a shadow. A vertex $s$ is in the shadow iff the straight line from parent $p$ to vertex $s$ passes through the interior of blocked cell $b$. We distinguish two parts of the perimeter of this shadow, namely the upper and lower boundary, as shown in Figure 25. We define a boundary vertex to be any vertex not in the shadow that has at least one neighbor (although not necessarily a visible neighbor) in the shadow. The origin $b_0$ is not in the shadow but its neighbor $south(b_0)$ is in the shadow. Thus, the origin $b_0$ is a boundary vertex. We consider only the upper boundary without loss of generality. Then, a boundary vertex (to be precise: an upper boundary vertex) is any vertex $s$ with $\Theta(s, p, b_0) \leq 0$ (that is, on or above the upper boundary and thus outside of the shadow) that has at least one neighbor $s'$ with $\Theta(s', p, b_0) > 0$ (that is, below the upper boundary and thus inside of the shadow). It is easy to see that all boundary vertices are in quadrant IV and form an infinite boundary path $[b_0, b_1, \ldots]$ that starts at the origin $b_0$ and repeatedly moves either south or east, that is, $b_{i+1} = south(b_i)$ or $b_{i+1} = east(b_i)$.

We define a vertex $s$ to be sufficiently constrained iff $\Theta(s, p, b_0) \leq lb(s)$ for its parent $p$. Once vertex $s$ is sufficiently constrained, it remains sufficiently constrained since no operation of AP Theta* can decrease its lower angle bound $lb(s)$. We prove in the following that every boundary vertex is sufficiently constrained at the time it is expanded if it is expanded with parent $p$. Consider any vertex $s$ below the upper boundary (that is, $\Theta(s, p, b_0) > 0$ and thus $\Theta(b_0, p, s) < 0$) that is a visible neighbor of some boundary vertex $b_i$. Vertex $s$ cannot have been updated according to Path 1 and been assigned parent $p$ at the time its parent $p$ was expanded since the straight line from parent $p$ to vertex $s$ passes through the interior of a blocked cell and they are therefore not visible neighbors. It cannot have been updated according to Path 2 and been assigned parent $p$ at the time boundary vertex $b_i$ was expanded with parent $p$ because boundary vertex $b_i$ is sufficiently constrained at that time and thus $\Theta(b_i, p, b_0) \leq lb(b_i)$, which implies that $\Theta(b_i, p, s) = \Theta(b_i, p, b_0) + \Theta(b_0, p, s) < \Theta(b_i, p, b_0) \leq lb(b_i)$ and the condition on Line 60 remains unsatisfied. Consequently, no vertex in the shadow can have parent $p$.

We now prove by induction on the order of the vertex expansions that every boundary vertex is sufficiently constrained at the time it is expanded if it is expanded with parent $p$. Assume that boundary vertex $b_0$ is expanded with parent $p$. Then, the condition on Line 81 is satisfied and Line 83 is executed for blocked cell $b$ at the time boundary vertex $b_0$ is expanded with parent $p$. Boundary vertex $b_0$ is sufficiently constrained afterwards since its lower angle bound is set to zero. Now assume that boundary vertex $b_i$ with $i > 0$ is expanded with parent $p$. Then, boundary vertex $b_i$ cannot be identical to parent $p$ (since they are in different quadrants) nor to the start vertex (since the start vertex does not have parent $p$). Boundary vertex $b_i$ cannot have been updated according to Path 1 and been assigned parent $p$ at the time its parent $p$ was expanded since $p.x < 0$ and $(b_i).x > 0$ and they are thus not neighbors. Consequently, boundary vertex $b_i$ must have been updated according to Path 2 and been assigned parent $p$ at the time one of its visible neighbors $x$ was expanded with parent $p$. Vertex $x$ must be on or above the upper boundary (that is, $\Theta(x, p, b_0) \leq 0$) and cannot be identical to parent $p$ (since they are in different quadrants). We distinguish two cases:

- Assume that vertex $x$ is a boundary vertex. It is sufficiently constrained at the time it is expanded with parent $p$ according to the induction assumption (that is, $\Theta(x, p, b_0) \leq lb(x)$) since it is expanded before boundary vertex $b_i$. Boundary vertex $b_i$ was updated according to Path 2 at the time vertex $x$ was expanded with parent $p$. Thus, the condition on Line 60 is satisfied at that time (that is, $lb(x) \leq \Theta(x, p, b_i)$) and thus $lb(x) + \Theta(b_i, p, x) = lb(x) - \Theta(x, p, b_i) \leq 0$. Then, the conditions on Lines 88 and 89 are satisfied and Line 90 is executed with $s' = x$ at the time boundary vertex $b_i$ is expanded with parent $p$. Boundary vertex $b_i$ is sufficiently constrained afterwards since its lower angle bound is set to $\max(lb(b_i), lb(x) + \Theta(b_i, p, x))$ and $\Theta(b_i, p, b_0) = \Theta(b_i, p, x) + \Theta(x, p, b_0) \leq lb(x) + \Theta(b_i, p, x) \leq \max(lb(b_i), lb(x) + \Theta(b_i, p, x))$.

- Assume that vertex $x$ is not a boundary vertex.

**Lemma 3.** *Assume that a vertex $s$ and a boundary vertex $b_i$ are visible neighbors, $c(p, b_i) < c(p, s)$ and $\Theta(s, p, b_i) < 0$. Assume that boundary vertex $b_i$ is sufficiently constrained at the time vertex $s$ is expanded with parent $p$ if boundary vertex $b_i$ has been expanded with parent $p$ at that time. Then, vertex $s$ is sufficiently constrained at the time it is expanded if it is expanded with parent $p$.*

*Proof.* Assume that vertex $s$ is expanded with parent $p$. Then, $\Theta(s, p, b_0) = \Theta(s, p, b_i) + \Theta(b_i, p, b_0) < 0$ since $\Theta(s, p, b_i) < 0$ and $\Theta(b_i, p, b_0) \leq 0$. We distinguish two cases:

- Assume that boundary vertex $b_i$ is not expanded before vertex $s$ or is expanded with a parent other than parent $p$. Then, the conditions on Lines 93 and 94 are satisfied and Line 95 is executed with $s' = b_i$ at the time vertex $s$ is expanded with parent $p$. Vertex $s$ is sufficiently constrained afterwards since its lower angle bound is set to $\max(lb(s), \Theta(s, p, b_i))$ and $\Theta(s, p, b_0) = \Theta(s, p, b_i) + \Theta(b_i, p, b_0) \leq \Theta(s, p, b_i) \leq \max(lb(s), \Theta(s, p, b_i))$.

- Assume that boundary vertex $b_i$ is expanded with parent $p$ before vertex $s$ is expanded with parent $p$. Boundary vertex $b_i$ is sufficiently constrained at the time vertex $s$ is expanded with parent $p$ according to the premise (that is, $\Theta(b_i, p, b_0) \leq lb(b_i)$). Furthermore, $lb(b_i) \leq 0$ (since no operation of AP Theta* can make the lower angle bound positive) and thus $lb(b_i) + \Theta(s, p, b_i) \leq 0$. Then, the conditions on Lines 88 and 89 are satisfied and Line 90 is executed with $s' = b_i$ at the time vertex $s$ is expanded with parent $p$. Vertex $s$ is sufficiently constrained afterwards since its lower angle bound is set to $\max(lb(s), lb(b_i) + \Theta(s, p, b_i))$ and $\Theta(s, p, b_0) = \Theta(s, p, b_i) + \Theta(b_i, p, b_0) \leq lb(b_i) + \Theta(s, p, b_i) \leq \max(lb(s), lb(b_i) + \Theta(s, p, b_i))$. $\qquad\qquad\square$

Boundary vertex $b_i$ is either immediately south or east of boundary vertex $b_{i-1}$ since the boundary path moves only south or east. We distinguish three subcases:

- Assume that parent $p$ is on the x-axis in quadrant II. Then, the boundary path is along the x-axis. Vertices $west(b_i)$ and $east(b_i)$ are boundary vertices, and vertices $southwest(b_i)$, $south(b_i)$, and $southeast(b_i)$ are below the upper boundary. Thus, vertex $x$ is identical to one of vertices $northwest(b_i)$, $north(b_i)$ or $northeast(b_i)$. In all cases, there is a boundary vertex $b_j$ immediately south of vertex $x$. If vertices $x$ and $b_j$ were not visible neighbors, then there would be blocked cells immediately southwest and south-east of vertex $x$ and vertices $x$ and $b_i$ could thus not be visible neighbors. Thus, vertices $x$ and $b_j$ are visible neighbors. Furthermore, boundary vertex $b_j$ is immediately south of vertex $x$ and thus $c(p, b_j) < c(p, x)$ and $\Theta(x, p, b_j) < 0$. Finally, boundary vertex $b_j$ is sufficiently constrained according to the induction assumption at the time boundary vertex $b_i$ is expanded with parent $p$ if boundary vertex $b_j$ has been expanded with parent $p$ at that time. Thus, vertex $x$ is sufficiently constrained at the time it is expanded with parent $p$ according to Lemma 3 (that is, $\Theta(x, p, b_0) \leq lb(x)$). Consequently, the conditions on Lines 88 and 89 are satisfied (for the reason given before) and Line 90 is executed with $s' = x$ at the time boundary vertex $b_i$ is expanded with parent $p$. Boundary vertex $b_i$ is sufficiently constrained afterwards since its lower angle bound is set to $\max(lb(b_i), lb(x) + \Theta(b_i, p, x))$ and $\Theta(b_i, p, b_0) = \Theta(b_i, p, x) + \Theta(x, p, b_0) \leq lb(x) + \Theta(b_i, p, x) \leq \max(lb(b_i), lb(x) + \Theta(b_i, p, x))$.

- Assume that parent $p$ is not on the x-axis in quadrant II and that boundary vertex $b_i$ is immediately east of boundary vertex $b_{i-1}$ and thus $c(p, b_{i-1}) < c(p, b_i)$ and $\Theta(b_i, p, b_{i-1}) < 0$. Furthermore, boundary vertex $b_{i-1}$ is sufficiently constrained according to the induction assumption at the time boundary vertex $b_i$ is expanded with parent $p$ if boundary vertex $b_{i-1}$ has been expanded with parent $p$ at that time. If boundary vertices $b_{i-1}$ and $b_i$ are visible neighbors, then boundary vertex $b_i$ is sufficiently constrained at the time it is expanded with parent $p$ according to Lemma 3. If boundary vertices $b_{i-1}$ and $b_i$ are not visible neighbors, then there must be blocked cells immediately north-west and south-west of boundary vertex $b_i$. Then, Line 81 is satisfied and Line 83 is executed for the blocked cell immediately south-west of boundary vertex $b_i$ at the time boundary vertex $b_i$ is expanded with parent $p$. Boundary vertex $b_i$ is sufficiently constrained afterwards since its lower angle bound is set to zero.

- Assume that parent $p$ is not on the x-axis in quadrant II and that boundary vertex $b_i$ is immediately south of boundary vertex $b_{i-1}$.

**Lemma 4.** *Assume that a vertex $s$ in quadrant IV is on or above the upper boundary. Then, vertex $s$ is a boundary vertex iff the vertex immediately south-west of vertex $s$ is below the upper boundary.*

*Proof.* If the vertex $s'$ immediately south-west of vertex $s$ is below the upper boundary, then vertex $s$ is a boundary vertex by definition. On the other hand, if vertex $s'$ is on or above the upper boundary (that is, $\Theta(s', p, b_0) \leq 0$), then vertex $s$ is not a boundary vertex because every neighbor of it is on or above the upper boundary. The neighbors of vertex $s$ are

$$east(s), northeast(s), north(s), northwest(s),$$
$$west(s), southwest(s), south(s) \text{ and } southeast(s).$$

or, equivalently,

$$east(east(north(s'))), east(east(north(north(s')))), east(north(north(s'))),$$
$$north(north(s')), north(s'), s', east(s') \text{ and } east(east(s')).$$

Thus, every neighbor $s''$ of vertex $s$ can be reached from vertex $s'$ by repeatedly moving either north or east and thus $\Theta(s'', p, s') \leq 0$. Consequently, $\Theta(s'', p, b_0) = \Theta(s'', p, s') + \Theta(s', p, b_0) \leq 0$ and thus every neighbor $s''$ of vertex $s$ is on or above the upper boundary. □

We distinguish two subcases:

* Assume that boundary vertex $b_{i+1}$ is immediately east of boundary vertex $b_i$. Vertices $north(b_i)$ and $east(b_i)$ are boundary vertices. Vertices $west(b_i)$, $southwest(b_i)$ and $south(b_i)$ are south-west of boundary vertices $b_{i-1}$, $b_i$ and $b_{i+1}$, respectively, and thus below the upper boundary according to Lemma 4. Vertices $northwest(b_i)$ and $southeast(b_i)$ are either boundary vertices or south-west of boundary vertices $b_{i-2}$ and $b_{i+2}$, respectively, and then below the upper boundary according to Lemma 4. Thus, vertex $x$ is identical to vertex $northwest(b_i)$.

* Assume that boundary vertex $b_{i+1}$ is immediately south of boundary vertex $b_i$. Vertices $north(b_i)$ and $south(b_i)$ are boundary vertices. Vertices $west(b_i)$ and $southwest(b_i)$ are south-west of boundary vertices $b_{i-1}$ and $b_i$, respectively, and thus below the upper boundary according to Lemma 4. Vertex $northwest(b_i)$ is either a boundary vertex or south-west of boundary vertex $b_{i-2}$ and then below the upper boundary according to Lemma 4. Thus, vertex $x$ is identical to one of vertices $northeast(b_i)$, $east(b_i)$ or $southeast(b_i)$.

In all cases, vertex $x$ is immediately east of some boundary vertex $b_j$ and thus $c(p, b_j) < c(p, x)$ and $\Theta(x, p, b_j) < 0$. If vertices $x$ and $b_j$ were not visible neighbors, then there would be blocked cells immediately north-west and south-west of vertex $x$ and vertices $x$ and $b_i$ could not be visible neighbors. Thus, vertices $x$ and $b_j$ are visible neighbors. Furthermore, boundary vertex $b_j$ is sufficiently constrained according to the induction assumption at the time boundary vertex $b_i$ is expanded with parent $p$ if boundary vertex $b_j$ has been expanded with parent $p$ at that time. Thus, vertex $x$ is sufficiently constrained at the time it is expanded with parent $p$ according to Lemma 3 (that is, $\Theta(x, p, b_0) \leq lb(x)$). Consequently, the conditions on Lines 88 and 89 are satisfied (for the reason given before) and Line 90 is executed with $s' = x$ at the time boundary vertex $b_i$ is expanded with parent $p$. Boundary vertex $b_i$ is sufficiently constrained afterwards since its lower angle bound is set to $\max(lb(b_i), lb(x) + \Theta(b_i, p, x))$ and $\Theta(b_i, p, b_0) = \Theta(b_i, p, x) + \Theta(x, p, b_0) \leq lb(x) + \Theta(b_i, p, x) \leq \max(lb(b_i), lb(x) + \Theta(b_i, p, x))$.

This concludes the proof that every boundary vertex is sufficiently constrained at the time it is expanded if it is expanded with parent $p$ and thus also the proof that AP Theta* never returns a path with a path segment that passes through the interior of a blocked cell.

We now prove that AP Theta* never returns a path with a path segment that passes between two blocked cells that share an edge. We prove by contradiction that AP Theta* cannot assign some parent $p$ to some vertex $s$ such that the path segment from parent $p$ to vertex $s$ passes between two blocked cells that share an edge. Assume otherwise and consider the first time AP Theta* assigns some parent $p$ to some vertex $s$ such that the path segment from parent $p$ to vertex $s$ passes between two blocked cells that share an edge. The path segment must be either horizontal or vertical. Vertex $s$ cannot have been updated according to Path 1 and been assigned parent $p$ at the time its parent $p$ was expanded since then the straight line from parent $p$ to vertex $s$ passes through the interior of a blocked cell and they are therefore not visible neighbors. It cannot have been updated according to Path 2 and been assigned parent $p$ at the time some visible neighbor $s'$ was expanded with parent $p$ since then either a) neighbor $s'$ would not be colinear with vertices $p$ and $s$ and the straight line from parent $p$ to vertex $s$ would thus pass through the interior of a blocked cell or b) neighbor $s'$ would be colinear with vertices $p$ and $s$ and the straight line from parent $p$ to vertex $s'$ would pass between two blocked cells that share an edge, which is a contradiction of the assumption. This concludes the proof that AP Theta* never returns a path with a path segment that passes between two blocked cells that share an edge.

Thus, AP Theta* never returns a blocked path. $\qquad\qquad\qquad\qquad\qquad\qquad\qquad\qquad\square$

## Appendix C. Acknowledgments

This article is an extension of an earlier publication (Nash et al., 2007) and contains additional expositions, examples and proofs. We thank Vadim Bulitko from the University of Alberta for making maps from the real-time game Baldur's Gate II available to us. Our research was done while Ariel Felner spent his sabbatical at the University of Southern California, visiting Sven Koenig. This research has been partly supported by a U.S. Army Research Laboratory (ARL) and U.S. Army Research Office (ARO) award to Sven Koenig under grant W911NF-08-1-0468, by a Office of Naval Research (ONR) award to Sven Koenig under grant N00014-09-1-1031, by a National Science Foundation (NSF) award to Sven Koenig under grant 0413196 and by an Israeli Science Foundation (ISF) award to Ariel Felner under grants 728/06 and 305/09. Alex Nash was funded by the Northrop Grumman Corporation. The views and conclusions contained in this document are those of the authors and should not be interpreted as representing the official policies, either expressed or implied, of the sponsoring organizations, agencies, companies or the U.S. government.